\documentclass[10pt,letterpaper]{article}

\usepackage[english]{babel}
\usepackage[utf8]{inputenc}
\usepackage[T1]{fontenc}
\usepackage{graphicx}
\usepackage{mdframed}
\usepackage[title]{appendix}

\usepackage{interval}
\usepackage{array}
\usepackage{booktabs}
\usepackage{colortbl}
\usepackage{comment}
\usepackage{subcaption}
\usepackage{amsmath}
\usepackage{algorithm}
\usepackage{algpseudocode}
\usepackage{multirow}
\usepackage{amsfonts}

\usepackage{relsize}
\usepackage{caption}
\usepackage{xcolor}
\usepackage{url}
\usepackage{cite}

\usepackage[a4paper,top=3cm,bottom=2cm,left=2cm,right=2cm,marginparwidth=1.75cm]{geometry}

\title{Synthetic flow-based cryptomining attack generation\\ through Generative Adversarial Networks}

\usepackage{authblk}

\author[1,*]{Alberto~Mozo}
\author[2,3]{\'Angel~Gonz\'alez-Prieto}
\author[4]{Antonio~Pastor}
\author[1]{Sandra~Gómez-Canaval}
\author[1]{Edgar~Talavera}
\affil[1]{Universidad Politécnica de Madrid, Madrid, Spain.}
\affil[2]{Universidad Complutense de Madrid, Madrid, Spain.}
\affil[3]{Instituto de Ciencias Matem\'aticas (CSIC-UAM-UCM-UC3M), Madrid, Spain.}
\affil[4]{Telefónica I+D, Madrid, Spain}
\affil[*]{a.mozo@upm.es}

\date{}

\begin{document}

\maketitle

\begin{abstract}
Due to the growing rise of cyber attacks in the Internet, flow-based data sets are crucial to increase the performance of the Machine Learning (ML) components that run in network-based intrusion detection systems (IDS).
To overcome the existing network traffic data shortage in attack analysis, recent works propose Generative Adversarial Networks (GANs) for synthetic flow-based network traffic generation.
Data  privacy  is  appearing  more  and  more  as a strong requirement when processing such network data, which suggests to find solutions where synthetic data can fully replace real data. 
Because of the ill-convergence of the GAN training, none of the existing solutions can generate high-quality fully synthetic data that can totally substitute real data in the training of IDS ML components. Therefore, they mix real with synthetic data, which acts only as data augmentation components, leading to privacy breaches as real data is used.

In sharp contrast, in this work we propose a novel deterministic way to measure the quality of the synthetic data produced by a GAN both with respect to the real data and to its performance when used for ML tasks.  As a byproduct, we present a heuristic that uses these metrics for selecting the best performing generator during GAN training, leading to a stopping criterion. An additional heuristic is proposed to select the best performing GANs when different types of synthetic data are to be used in the same ML task.
We demonstrate the adequacy of our proposal by generating synthetic cryptomining attack traffic and normal traffic flow-based data using an enhanced version of a Wasserstein GAN. 
We show that the generated synthetic network traffic can completely replace real data when training a ML-based cryptomining detector, obtaining similar performance and avoiding privacy violations, since real data is not used in the training of the ML-based detector.
\\

\end{abstract}

\section*{Introduction}

Cybersecurity and large-scale network traffic analysis are two important areas receiving considerable attention over the last few years. Among other reasons, this is due to the necessity of empowering the telecom industry to adopt suitable mechanisms to face emerging and sophisticated cyberattacks.
Nowadays, Internet Service Providers (ISPs) and their clients are exposed to a growing rise in the number and type of threats (e.g., network attacks, data theft over the wire), some of which also attack at the application level using the network for identity theft, phishing, or malware distribution. 
In general terms, these threats severely put QoE (Quality of Experience) at risk, undermining services, network resources, and users' confidence.
In this context, one promising solution is the use of Machine and Deep Learning (MDL) techniques to address the appearance of new points of vulnerability and exposure to new attack vectors \cite{ENISA, dasgupta2020machine, mahdavifar2019application}. 
At the same time, malicious agents are moving forward in the same direction to use MDL for their activities or to deceive MDL inference engines \cite{EUROPOL}.

The application of MDL techniques requires the availability of considerable amounts of data to take advantage of their powerful learning processes. 
Telecom data management processes are not well suited to offer these required data sets as they exhibit a set of problems not only are related with the gathering and sharing of data but also with their processing in a MDL pipeline.
This situation represents a considerable drawback since  data gathering and processing tasks in the telecom industry have been optimized to guarantee services and billing. Indeed, they are not prepared with specific MDL data processing techniques.
Moreover,
the applicability of MDL algorithms should take into account the evolution of attack patterns over time, which implies to produce periodically additional  volumes of relevant data for training new MDL models.

Moreover, a great percentage of MDL techniques used in Intrusion Detection Systems (IDS) are the so-called supervised techniques that require labelled data sets to train and validate MDL models. As in many other domains, telecom industry faces the impossibility of having labelled data sets or developing efficient and accurate processes to label them.
Since network traffic is generated by end users and applications, it can be challenging for an ISP to identify and label the nature of network traffic at the detailed level required by MDL techniques.
This difficulty is exploited by cyber criminals, who seek to mix cyber attacks with normal traffic by encrypting it over common TCP ports (e.g., Transport Layer Security (TLS) using TCP/443 (HTTPS)).
Although unsupervised techniques that do not need labelled data sets can be applied in some scenarios, a significant number of sophisticated attacks require supervised MDL methods to be detected. 

Even if efficient mechanisms for labelling data sets can be implemented, data are increasingly protected by the legal regulations that governments impose to guarantee the privacy of their contents (e.g., European General Data Protection Regulation (GDPR)). These restrictions may discourage the use of real data sets for MDL training and validation purposes.
For a suitable advance on cybersecurity research, and specifically, on threat detection in network traffic, the telecom industry 
requires novel methods to generate labelled data sets to be used in MDL training and validation processes.

In the last decade, Generative Adversarial Networks (GANs) \cite{Goodfellow:2014} have gained significant attention due to its ability to generate synthetic data simulating realistic media such as images, text, audio and videos \cite{wang2019generative, gao2020generative, jabbar2020survey,pan2019recent}. 
Nowadays, GANs are broadly studied and applied through academic and industrial research in different domains beyond media (e.g., natural language processing, medicine, electronics, networking, and cybersecurity).
In short, a GAN model is represented by two independent neural networks (the generator and the discriminator) that compete to learn and reproduce the distribution of a real data set. 
After a GAN has been trained, its generator can produce as many synthetic examples as necessary, providing an efficient mechanism for solving the lack of labelled data sets and potential privacy restrictions. 

In this context, this work proposes the application of GANs to generate synthetic flow-based network traffic that mimicks cryptomining attacks and normal traffic.
In contrast to most of the proposed works that are based on data augmentation solutions, we aim to generate synthetic data that can fully replace real data (attacks and normal traffic). Therefore, MDL models trained with synthetic data will obtain a similar performance to MDL models trained with real data when both are tested and deployed in real-time scenarios.

This solution has two clear advantages: Firstly, addressing the existing shortage of publicly available network traffic datasets containing attacks and normal traffic and secondly, avoiding the privacy violations that could appear when real data is used in MDL training and testing processes.

In the light of these advantages, interesting applications can be devised. The first one is related to MDL cross-developments. Providing labelled data that does not incur in privacy breaches can foster cross-development of MDL components by third parties. For example, a telecom provider  developing ML-based components to be part of an IDS, receives synthetic data from a telecom operator to train and validate these ML-based components. As the synthetic data have been generated from real data using GANs, the ML component after training will reach the desired level of performance and furthermore, no breach of data privacy  will be raised as the telecom operator is not sharing any real data with the telecom provider.

In addition, the solution proposed in this work is useful for application in Cyber-range exercises. Cyber ranges are  well  defined  controlled  virtual  environments  used  in  cybersecurity  training  as an efficient  way  for  trainees  (e.g. cyber-security personnel) to  gain  practical  knowledge  through  hands  on  activities \cite{neville2009rational, ferguson2014national}.

Synthetic flow-based network traffic and attacks generated by GANs can be used in cyber ranges to generate different data for a concrete type of exercise and avoid blue teams learning such exercise always with the same data. 
Having trained a GAN model to replicate a given type of attack (or normal traffic), we can generate as many attacks of such type as required. Therefore, even if the blue team repeats an exercise several times, the analised attacks and normal traffic are not going to be exactly the same in each run of the exercise. 
In addition, red teams can use GANs in penetration testing (pentest) exercises to generate realistic attacks that never contain the same attack data even if the launched attacks are of the same type.
Thus, the robustness of an IDS against a type of attack can be  evaluated launching many different synthetic samples of the same attack. 

Furthermore, a Cyber-range can import from third parties data sets containing attacks and normal traffic that are subject to privacy or anonymity restrictions. As the network  data used in the exercises by the blue and red teams are the synthetic ones generated by the GANs, no breach of privacy appears during the realization of such exercises. Moreover, exporting attacks and normal data (e.g. to other platforms in a federated cyber range) can be done without incurring in any privacy violation as the exported data to be shared with a third entity are exclusively the synthetic network traffic generated by the GANs.

These ideas have been applied by the authors of the manuscript in the H2020 SPIDER project
\cite{SPIDER_2020} that proposes a cyber-range solution that is specifically designed to train cybersecurity experts in telecommunications environments and covers all cybersecurity situations in a 5G network environment. 
As a novelty in cyber-ranges, SPIDER brings a way to seamlessly integrate ML-based components and GANs to be used as part of blue and red team toolboxes.
The GAN models proposed in this manuscript will be used in SPIDER as the basic building block of the Synthetic Network Traffic Generator to obtain synthetic network traffic data (attacks and well-behaved connections) that reproduce the statistical distribution of real traffic to be used later in cyber range exercises.

\subsection*{Proposal}
To demonstrate the applicability of our proposal, we select a cryptomining attack scenario. Cryptomining is a paradigmatic cryptojacking attack that is gaining momentum in these days. 
Cryptomining attacks concern the network traffic generated by cybercriminals that create tailored and illegal processes for catching computational resources from users’ devices without their consent to use them in the benefit of the criminal for mining cryptocurrencies. It has been shown that these malicious connections can be detected in real-time with decent accuracy  even at the very beginning of the connection's lifetime by using an ML classifier \cite{pastor2020detection}.

Our goal is to obtain WGAN synthetic traffic of sufficient quality to allow a complete replacement of real data by synthetic samples during the training of a ML-based crytomining attack detector.  This  property ensures that we will not violate any privacy restriction and sets our proposal apart from existing works that only propose data augmentation solutions based on mixing real with synthetic data.

To generate flow-based information that replicates normal traffic and cryptomining connections, we apply two Wasserstein GANs \cite{Arjovsky-WGAN} to generate both types of network traffic separately.
Unlike current solutions, our WGANs replicate not only already completed connections, but also connections at different stages of their lifetime.
Regarding that successful GAN training is still an open research problem \cite{bengio2021deep}, we propose to evaluate a set of GAN enhancements to measure their impact on the convergence of the WGAN training and the quality of the synthetic data generated. 

In addition, we propose two new metrics based on the $L^1$ distance and the Jaccard coefficient to measure the quality of the synthetic data generated by GANs with respect to their similarity with real data. 
These new metrics consider the joint distribution of all variables of the flow-based data rather than the mean over the distance of each variable as other works propose.

To the best of our knowledge and due to the ill-convergence of GANs, none of the existing works propose a clear stopping criterion during GAN training to obtain the best performing synthetic data. 
To address this problem, we propose a simple heuristic for selecting the best performing GANs when it is required to fully replace real data with synthetic data for training MDL models. 
This heuristic exploits our experimental observations in three aspects: 
\begin{enumerate}
    \item When synthetic data is used for training a MDL model, its performance is not well correlated with any quality metric that can be applied to the synthetic data, and therefore, it is required to  train and validate a MDL model with a sample of the synthetic data to obtain the ML performance (e.g. $F_1$-score on testing).
    \item After having trained thousands of GANs models, we observed that the performance value we used ($F_1$-score) tends to oscillate during training, and therefore, stopping training because the cost function is not decreasing or the quality of the data is not improving is not an appropriate criterion as performance could improve after additional training. In order to obtain the best performing GAN we suggest to measure GAN performance at the end of each mini-bath training.
    \item Even if we select the best performing GAN for each type of traffic, it is not guaranteed that their joint performance will be the best when both types of synthetic data are combined to fully replace the real data during the training of MDL models.  
\end{enumerate}

The first and second observations involve evaluating the performance of each generator obtained at the end of a mini-batch training stage to select the best one. The third observation indicates that even when selecting the best performing generator for each type of traffic, the performance obtained when we mix them may not be the best, and therefore we propose a heuristic for finding the best performing combination. A random selection of the intermediate GAN models obtained during training works well, requiring only a moderate number of samples. Nevertheless, we observed that when $F1-score$ on testing was used as the performance metric, reducing the sample GAN universe to the top-10 best performing for each type of traffic produced similar results requiring a significantly fewer number of samples.  

\subsection*{Experiments}

To demonstrate the proposed solution, we ran an extensive set of experiments. 
The data sets used in our experiments were previously generated in a realistic network digital twin called the Mouseworld lab \cite{pastor2018mouseworld, mozo2018distributed}. In this lab, we launched real clients and servers  that interacted with other hosts located in different places in the Internet and collected the generated traffic composed of encrypted and non-encrypted flows (normal traffic and cryptomining connections). 
A set of 59 statistical features were extracted from each TCP connection in real time each time a new packet was received. We carefully selected a reduced set of 4 features for our GAN experiments.

We trained independently two WGANs, one for each type of traffic, configuring them with a rich  set of hyperparameters.
We performed a blind random search in the hyperparameter space of each WGAN and to select the best configuration of hyperparameters we used the $F_1$-score obtained in a nested ML-model that was executed after each train epoch. 
For each type of traffic, we selected the WGAN that obtained the best $F_1$-score for the nested classifier in any of its epochs. 
In addition, we compute the two proposed metrics ($L^1$ distance and Jaccard) on each WGAN to analyse the similarity of the synthetic traffic they are generating with respect to the real data.
Given that for each winning WGAN, we have as many intermediate generators as the train steps we run, the previously proposed heuristic is run to choose pairs of generators that produce good results in a global nested evaluation of both WGANs. Thus, we avoid testing all possible combinations of generators from the two WGANs.  

To measure the quality of the synthetic data generated by our GANs, we train a ML classifier for detecting cryptomining connections in real-time using only a combination of the synthetic traffic generated by the two WGANs (normal traffic and cryptomining attacks). Another ML classifier configured with the same set of hyperparameters is trained using real data. Both models are tested against a second set of real data to measure whether the ML model trained exclusively with synthetic data performs at the same level than the model trained with real data.
As a baseline for the quality of synthetic traffic, we use a naive approach based on a noise generator added to the averages of the variables and as an upper bound, we consider the results obtained with real traffic.
In addition, we run the same experiments but adding some extensions to the standard WGAN configurations to analise whether the performance of the nested ML or the convergence of the training process increase when these heuristics are considered.

\subsection*{Paper structure}
The rest of the paper is structured as follows: Section \ref{sec:related_work} presents the related work in this topic and the manuscript contributions are detailed in  subsection \ref{subsec:contribution}.
The problem setting is shown in Section \ref{sec:problem-setting}. 
The proposed model is depicted in Section \ref{sec:proposed_model} and the model architecture, generator custom activation functions and WGAN enhancements and heuristics are explained in specific subsections. The proposed quality and similarity metrics are detailed in Section \ref{sec:metrics}. The empirical evaluation using standard WGANs is presented in Section \ref{sec:empirical} and the effects of several improvements and variants are shown in Section \ref{sec:improvements}.
We conclude and summarize with some interesting open research questions in Section \ref{sec:conclusions_future-work}.

\section{Related work}
\label{sec:related_work}

Over the past, a few different research works have targeted the generation of synthetic network traffic using GANs\cite{navidan2021generative}, although the majority of them only propose data augmentation solutions  that are not applicable in scenarios in which data privacy must be guaranteed as they use a combination of real and synthetic data.

A recent work \cite{Rigaki2018} proposes a GAN approach to generate network traffic that mimics the traffic from Facebook's chat to modify the behavior of a real malware imitating  a normal traffic profile based on input parameters from the GAN .
This approach is able to modify malware communication in order to avoid detection. 
MalGAN \cite{Hu2017} is a GAN that can generate synthetic adversarial malware examples, which are able to bypass black-box machine learning based detection models. 
Bot-GAN \cite{Yin2018} presents a GAN-based framework to augment botnet detection models generating synthetic network data.
An augmentation method based on AC-GANs  was proposed to generate synthetic data samples for balancing network data sets containing only two classes of traffic \cite{Vu2017}. No details are given of either the structure of the GAN or how the experiments were evaluated, making it impossible to reproduce the proposed solution or measure the synthetic data quality.
Another similar work \cite{Wang2019} used GANs and the  ``ISCX VPN nonVPN" traffic dataset \cite{draper2016characterization}   but without proposing any evaluation method to contrast the obtained results.

A Deep Convolutional Generative Adversarial
Network (DCGAN) was recently proposed as a semisupervised data augmentation solution \cite{Iliyasu2020}. Samples generated by the DCGAN generator as well as unlabeled data are used to improve the performance of a basic CNN classifier trained on a few labeled samples from the  "ISCX VPN nonVPN" traffic dataset.
Another work \cite{Salem2018} presents a Cycle-GAN to augment and balance the ADFA-LD dataset containing system calls of small footprint attacks . Foot-prints are converted into images to be processed by Cycle-GANs in the standard way.  
A data augmentation method  using the NSL-KDD data set is proposed to generate adversarial attacks that can deceive and evade an intrusion detection system  \cite{Lin2018}. 
No details are provided on the network structure and hyperparameters used during training, which prevents experimental replicability. In addition, the proposed solution only uses flow statistics after the connection has finished, which restricts its applicability to forensic scenarios. 

PAC-GAN \cite{Cheng2019} proposes a different approach to generate packets instead of flow-based data. Authors assume that the bytes in a packet have a topological structure in order to apply convolutional neural networks in the generator. The generator implements 3 types of request packets (Ping request, HTTP Get and DNS request). Only request packets are generated because the quality metric applied only counts the number of responses received when these synthetic packets are sent to a server. Unfortunately, this metric cannot detect whether the GAN is learning the input packets by heart and simply replicating them at the output. The limited number of packet types that can be generated, together with the fact that they do not propose a realistic  metric for measuring the quality of the data generated, discourages the use of this solution in realistic environments.

A close work to our research is proposed in  \cite{Ring2019} where   three GAN variations are used  to generate flow-based network traffic information in unidirectional Netflow format. 
As GANs generate continuous values, this work presents a method inspired by Word2Vec from the NLP domain to transform  categorical variables (e.g., IP addresses) into continuous variables. 
The dataset used as input to the proposed GANs is a publicly available CIDDS-001 data set \cite{ring2017creation} that contains a mixture of normal traffic and attack connections. 
Unfortunately, only one record per connection is stored containing the final status of flow variables, which precludes its application in real-time IDS scenarios where flow-based information is needed to be generated recurrently along the life of the connection to detect malicious flows in their early beginning.
In addition, the authors propose to use as quality metrics the Euclidean distance between variables and manual techniques such as visual inspection and domain value checking done by experts that are impractical from a scalability perspective.
Moreover, the Euclidean distance measures each variable separately when what needs to be measured is the distance of the synthetic distribution from the real distribution but considering both as multivariate distributions. Finally, the authors do not propose as validation metric a performance evaluation of a ML-based attack detector trained with the synthetically generated flow-based data.

Cyber  ranges  are  well  defined  controlled  virtual  environments  used  in  cybersecurity  training  as an efficient  way  for  trainees  to  gain  practical  knowledge  through  hands  on  activities.
Several works have been proposed to study and classify the concept of a cyber range  covering different types of cyber ranges and security testbeds \cite{davis2013survey, yamin2020cyber, ukwandu2020review}.
Although in recent years Artificial Intelligence and Machine Learning based  technologies have been started to be actively used for cyber defense purposes \cite{yamin2021weaponized, kamoun2020ai, kaloudi2020ai}, their inclusion in cyber ranges is still in their infancy.

The H2020 SPIDER project \cite{SPIDER_2020}, in which we are applying the main results of this manuscript, proposes a cyber range solution, covering all cybersecurity situations in a 5G network environment, where ML-based components can be seamlessly integrated in blue and red team toolboxes.
The SPIDER framework is able to produce offensive and defensive artifacts using Synthetic Network Traffic Generators that can emulate specific types of attacks and normal traffic. 
To the best of our knowledge, only the H2020 SPIDER project is proposing to apply GANs to generate such synthetic attacks and normal traffic that can be used by red and blue teams. 
Thus, red teams can use the synthetic attacks to break the robustness of an IDS during pentest exercises and blue teams can learn how to reconfigure IDS defenses when faced with different synthetic attacks.
This novel feature circumvents the current limitations of cyber range commercial products that, using a fixed set of previously stored attacks, only generate slight variations of these attacks mainly based on adding noise combinations to the mean values.

\subsection{Contribution}
\label{subsec:contribution}
The main differences of the previous proposals and our work are: (i) Unlike the prevailing existing data augmentation solutions, we obtain synthetic flow-based traffic that can fully replace real data and therefore, this solution can be applied in scenarios where data privacy must be guaranteed,  (ii) existing metrics measure the differences of synthetic and real data independently for each each variable, but we propose a set of new metrics to measure more realistically the similarity of the two as joint multivariate distributions, (iii) due to the ill-convergence of the GAN training, none of the published papers mention a clear stopping criterion during training for selecting the best performing GAN, but we propose a simple heuristic that selects such GAN by measuring the performance of a ML task in which real data is fully replaced by  synthetic data for the task training, (iv) the proposed heuristic is extended to  efficiently  choose generators from different GANs to generate a combination of high quality synthetic data to be used in the same ML task -- in our scenario we generate a mixture of two different types of synthetic traffic to train a ML-based cryptomining attack detector--, 
(v) our solution does not generate one flow-based element at the end of the connection, but a set of elements representing different instants of a connection throughout its lifetime, which allows its usage and deployment in real-time scenarios, and (vi) we selected a recently appeared cryptomining attack as a paradigmatic use case to demonstrate the feasibility of our proposal and how it can be integrated into a next-generation cyber range.


\begin{figure}[!t]
    \centering
    \includegraphics[width=0.8\textwidth]{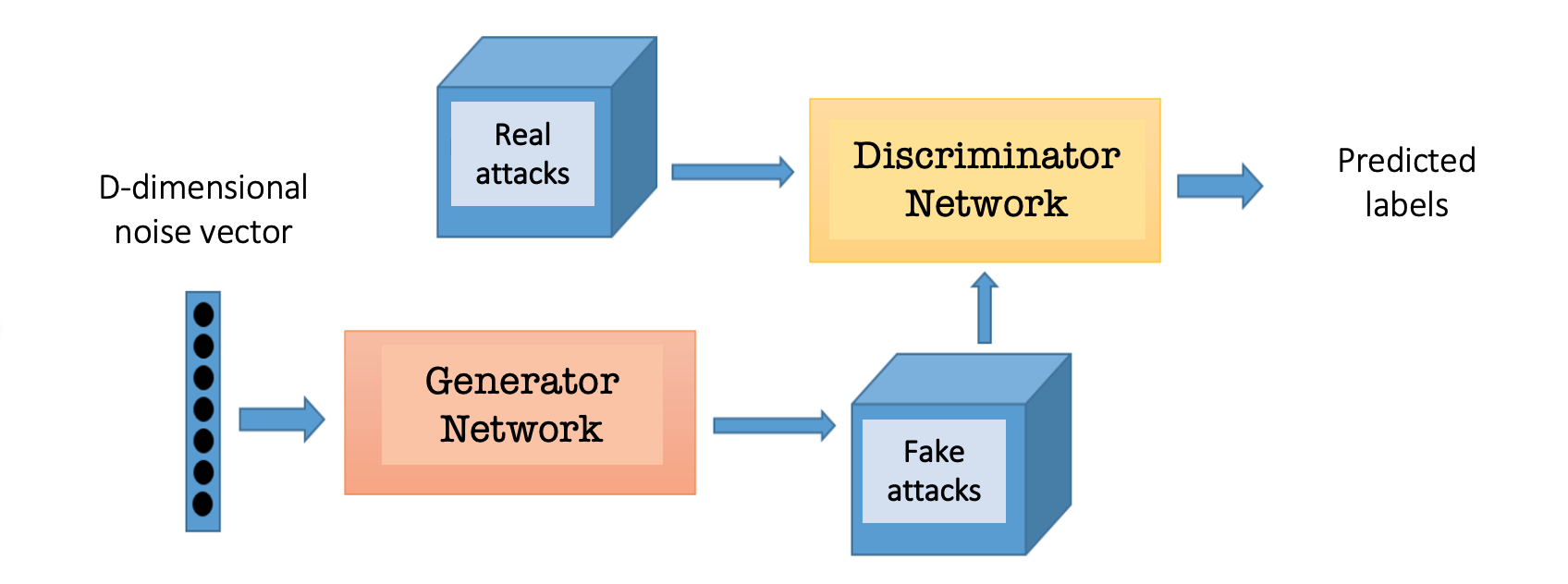}
    \caption{GAN architecture used as reference model.}
    \label{fig:aquitecture-gan}
\end{figure}

\section{Problem setting} 
\label{sec:problem-setting}

The framework of this work is a network environment in which real clients and servers compete  to exchange different types of traffic  sharing Internet connectivity. On one side, normal traffic clients interact with servers (web surfing, video and audio streaming, cloud storage, file transfer, email and P2P among others) but, on the other side, cryptomining clients connect to real mining pools generating a certain amount of cryptomining-related traffic that competes with the normal traffic for processing time and/or bandwidth. 
In this context, cybercriminals can populate their cryptocurrency wallets by using botnets or run illegal processes in browsers to surreptitiously add victims' computer resources  without their consent and spend computational resources for the criminal's benefit. 

For this reason, it is crucial to develop a system capable of identifying cryptomining-related traffic so that an attack detector can limit the bandwidth dedicated to this type of traffic or, in a extreme situation, block the connection. 
To be usable for this purpose, this detection must be conducted in real-time. 
Nowadays, some solutions for cryptomining detection are based on ML-based binary classifiers, such as random forest, decision trees or logistic regression. In order to prepare these models, it is necessary to feed them with lots of labelled data. With this data, during the so-called training process, an optimization algorithm adjusts the internal parameters of the model to extract the patterns that identify the cryptomining traffic with respect to the normal traffic so that this knowledge can later be used to classify the traffic.
Despite this is the customary procedure in ML, in this scenario this solution poses an important privacy problem. Indeed, the data needed for the training process is typically an excerpt of the real traffic crossing the node, for which a thorough offline analysis of its nature has to be conducted to identify the cryptomining traffic.

Regarding that companies are reluctant to share their data with third party developers and governments concerned about privacy issues are imposing limitations to telecom providers for accessing user data or inspecting packet payload,
the aim of this work is to construct a generative model able to substitute these data belonging to real clients by fully synthetic data with no information of the original clients. With this solution at hand, the Internet service providers are able to generate as much new anonymous data as needed to train their ML models or share it with third party developers 
without generating any privacy breach or jeopardizing the privacy of their users. 

The generated data should be as faithful to the real data as possible in the sense that they should provide significant information to the ML models to extract the underlying patterns distinguishing between normal and cryptomining traffic. In this sense, the performance of the generative models will be evaluated based on their ability to create high-quality synthetic data. Explicitly, the generated data must lead to a similar performance of the ML-based models when tested against models trained with real traffic, even though they were trained with synthetic data instead of real data.

As the original dataset to be replicated, we shall use a collection of flow-based statistics extracted from 4 hours of real traffic that was generated in a realistic network digital twin called the Mouseworld lab. This data was gathered in a controlled location of the Internet in two different instants of time: the first gathering will be used as the training dataset to be replicated and the second  will be stored as the testing dataset. The details of this process can be found in Section \ref{subsec:testbed}.

For each traffic flow (TCP connection), we compute a set of $59$ statistical variables describing the flow. The computation is carried out each time a packet for this connection is received or when a timeout fires. For our GAN experiments, 4 variables were selected: (a) number of bytes sent from the client, (b) average round-trip time observed from the server, (c) outbound bytes per packet, and (d) ratio of packets\_inbound / packets\_outbound.
It is worth noting that other sets of statistical variables are representatives of a flow (TCP connection) and may be computed. In fact, the Tstat tool \cite{tstat-finamore2011experiences} used for this task extracts and computes a total of $140$ variables. The four chosen features were selected as they exhibit several interesting properties for our generative experiments.
\begin{itemize}
    \item These four features themselves lead to a good performance in a standard ML classifier when used to classify between normal and criptomining traffic.
    \item Each feature exhibits a different statistical behaviour, which allows to demonstrate that the proposed solution can replicate a variety of data distributions and not only the normal one.
    \item The average value of each feature in the two types of traffic (normal and cryptomining) were close and, therefore, the traffic ML classifier needed to learn something subtler from the data features than their means. This property allows us to quantify the improvement obtained by using the proposed generative model versus a na\"ive Gaussian generator that produces data around a mean value. This is particularly interesting when we want to replicate data distributions that do not follow a normal distribution or have some hard domain constrains such as non-negativity or discrete distributions, such as those fulfilled by the chosen features.
\end{itemize}

\section{Proposed model}
\label{sec:proposed_model}

The solution we propose to address the problem described in Section \ref{sec:problem-setting} is based on Generative Adversarial Networks (GANs), as introduced by Goodfellow \cite{Goodfellow:2014}. A GAN network is a generative model in which two neural networks compete to improve their performance. To be precise, we have a $d$-dimensional random vector $X: \Omega \to \mathbb{R}^d$, defined on a certain probability space $\Omega$, that returns instances of a certain phenomenon that we would like to replicate. Usually, we have that $\Omega = \left\{x_1, \ldots, x_N\right\}$ is a large dataset of vectors $x_i \in \mathbb{R}^d$ and $X$ just picks randomly (typically uniformly) an element $x_i \in \Omega$. Moreover, in standard applications of GANs, we have that the instances $x_i$ are images represented by their pixel map and the objective of the GAN is to generate new images as similar as possible to the ones in the dataset.

For this purpose, a classical GAN proposes to put two neural networks to compete: a neural network $G$, called the generative network, and another neural network $D$, called the discriminant. The discriminant is a network computing a function $D: \mathbb{R}^d \to \mathbb{R}$ that is trained to solve a typical classification problem: given $x \in \mathbb{R}^d$, $D(x)$ is intended to predict whether $x = X(\omega)$ for some $\omega \in \Omega$ or not i.e.\ whether $x$ is compatible with being a real instance or it is a fake datum. Observe that, along this paper, we will follow the convention that $D(x)$ is the probability of being real, so $D(x)=1$ means that $D$ is sure that $x$ is real and $D(x) = 0$ means that $D$ is sure that $x$ is fake. On the other hand, the generative network computes a function $G: \mathbb{R}^l \to \mathbb{R}^d$. The idea of this function is that $\mathbb{R}^l$ will be endowed with a probability distribution $\lambda$, typically a spherical normal distribution or a uniform distribution on the unit cube. The probability space $\Lambda = (\mathbb{R}^l, \lambda)$ is called the latent space and the goal of the generator network is to tune $G$ in such a way that the random variable $G: \Lambda \to \mathbb{R}^d$ distributes as similar as possible to $X$.

The competition appears because the networks $D$ and $G$ try to improve non-simultaneously satifactible objectives. On the one hand, $D$ tries to improve its performance in the classification problem but, on the other hand, $G$ tries to generate as best results as possible to cheat $D$. To be precise, observe that the perfect result for the classification problem for $D$ is $D(x)=1$ is $x$ is an instance of $X$ and $D(x)=0$ if not. Hence, the mean error made by $D$ is
$$    \mathcal{E} = \mathbb{E}_\Omega \left[1-D(X)\right] + \mathbb{E}_\Lambda \left[D(G)\right] = 1 - \mathbb{E}_\Omega \left[D(X)\right] + \mathbb{E}_\Lambda \left[D(G)\right],
$$
where $\mathbb{E}_{\Omega}$ and $\mathbb{E}_{\Lambda}$ denote the mathematical expectation on $\Omega$ and $\Lambda$ respectively. In this way, the objective of $D$ is to minimize $\mathcal{E}$ while the objective of $G$ is to maximize it. It is customary in the literature to consider as objective the function $1-\mathcal{E}$ and to weight the error with a certain concave function $f: \mathbb{R} \to \mathbb{R}$. In this way, the final cost function is
$$
    \mathcal{F}(D,G) = \mathbb{E}_\Omega f\left[D(X)\right] + \mathbb{E}_\Lambda f\left[-D(G)\right] 
$$
and the objective of the game is
$$
    \min_{G}\,\max_{D} \mathcal{F}(D,G) = \min_{G}\,\max_{D}\;\mathbb{E}_\Omega f\left[D(X)\right] + \mathbb{E}_\Lambda f\left[-D(G(Z))\right].
$$
Typical choices for the weight function $f$ are $f(s) = - \log(1 + \exp(-s))$, as in the original paper of Goodfellow \cite{Goodfellow:2014}, or $f(s) = s$ as in the Wasserstein GAN (WGAN) \cite{Arjovsky-WGAN}. This operation method can be depicted schematically as in Figure \ref{fig:aquitecture-gan}.

Despite the simplicity of the formulation of the cost function, the optimization problem is far from being trivial. 
The best scenario would be to obtain a so-called Nash equilibrium for the game, that is, a pair of discriminant and generative networks $(D_0,G_0)$ such that the function $D \mapsto \mathcal{F}(D, G_0)$ has a local maximum at $D = D_0$ and the function $G \mapsto \mathcal{F}(D_0, G)$ has a local minimum at $G = G_0$. In other words, at a Nash equilibrium, neither $D$ or $G$ can improve their result unilaterally. Based on this idea, the classical training method as proposed by Goodfellow in \cite{Goodfellow:2014} is alternating optimization of $D$ and $G$ using classical gradient descend-based backpropagation. Despite that this method may provoke some convergence issues, as mentioned below, it is a widely used learning algorithm due to its simplicity and direct implementation using standard machine learning libraries like Keras or TensorFlow.

To be precise, the algorithm proposed by Goodfellow suggests to freeze the internal weights of $G$ and to use it to generate a batch of fake examples from $\Lambda$. With this set of fake instances and another batch of real instances created using $X$ (i.e.\ sampling randomly from the dataset of real instances), we train $D$ to improve its accuracy in the classification problem with the usual backpropagation (i.e.\ gradient descent) method. Afterwards, we freeze the weights of $D$ and we sample a batch of latent data of $\Lambda$ (i.e.\ we sample randomly noise using the latent distribution) and we use it to train $G$ using gradient descent for $G$ with objective $f(-D(G(z)))$. We can alternate this process as many times as needed until we reach the desired performance.

As noted in a recent work \cite{Nagarajan-Kolter}, the game to be optimized is not a convex-concave problem, so in general the convergence of the usual training methods is not guaranteed. 
Under some assumptions on the behaviour of the game around the Nash equilibrium points, it is proved that the usual gradient descent optimization is locally asymptotically stable \cite{Nagarajan-Kolter}. However, the hypotheses needed to apply this result are quite strong and seem to be unfeasible in practice. For instance, it has been published an example of a very simple GAN, the so-called Dirac GAN,  for which the usual gradient descend does no converge \cite{Mescheder-Geiger:convergence}.

For this reason, several heuristic methods for stabilizing the training of GANs have been proposed such as feature matching, minibatch discrimination, and semi-supervised training \cite{Salimans-Goodfellow:2016} as well as approaches changing the weight function $f$ as the Wasserstein GAN \cite{Arjovsky-WGAN}. The most promising approaches are based on the modification of the usual alternating gradient descending optimization such as the introduction of instance noise \cite{Sonderby:2017, Arjovsky-Bottou:2017} and regularization methods based on gradient penalty \cite{Roth-Lucchi}. A recent work \cite{gonzalez2021dynamics} proposes a formal study of the dynamics of the GAN training process, but due to the complexity of the analysis, two simplified neural network architectures and a torus space were considered. For a thorough analysis of the different methods for stabilizing the training of GANs, see \cite{Mescheder-Geiger:convergence}. 

The study of the type of network and the best architecture for $D$ and $G$ have been intensively studied in the literature. Convolutional neural networks for $D$ and deconvolutional networks for $G$ seem to be good choices for image generation and discrimination \cite{Radford-Metz}, as the set of variables (i.e pixels) in an image have topological information that can be exploited by these convolutional networks. However, due to their simplicity, fully connected NN (multilayer perceptrons) have also been applied to GANs with pretty good performance \cite{Goodfellow:2014}, in particular when no topological information is contained in the variables to be replicated.

\subsection{Architecture}
\label{subsec:architecture}

Aiming to mimic two types of behaviour (cryptomining attacks and well-behaved connections), in preliminary experiments, we adopted a well-known conditional GAN model, the so-called Auxiliary Classifier GANs (AC-GAN) \cite{odena2017conditional}, as the architecture to generate at the same time the two types of traffic variables. As this strategy did not produce an adequate performance when replicating the two types of traffic and generated significant oscillations in the convergence process, we opted to use a different approach. 
We conjecture that the oscillatory behaviour could be caused by the fact that cryptomining connections follow a very specific statistical pattern, and on the contrary, the normal traffic connections are made of a mixture of many different connections that globally exhibit a nearly random behaviour. In addition, there is a great imbalance in the number of data for each of the two types of connections, which might incline the GAN training to obtain one distribution closer to the real data than the other.

Assuming that the two types of traffic are independent each other and therefore it is not necessary to use one for the synthetic generation of the other, we finally proposed to train independently two standard GANs, one for normal traffic (i.e. well-behaved connections) and the other for the cryptomining connections. The reference architecture for these GANs is shown in Figure \ref{fig:aquitecture-gan}. As previously explained, this model is composed of two networks, a generator and a discriminator, competing between them.
The generator is input with a  random noise vector and produces a synthetic sample.  
The discriminator receives real and fake samples as input and tries to classify them appropriately in their correct category.
During training, the goal of the generator is to learn how to produce fake samples that can be classified as real by the discriminator. On the contrary, the goal of the discriminator is to learn how to differentiate real from fake examples.

To get rid of the mode collapse problems that frequently appear during GAN training, we adopted as a reference model the WGAN architecture \cite{Arjovsky-WGAN}, in which a  Wasserstein loss function is used as loss function instead of a standard cross-entropy function. A detailed explanation of why using W-GANs over standard GANs enhances the convergence during the GAN training process can be found in \cite{weng2019gan}. In addition to the replacing of the loss function, we tested two different strategies to enforce the required Lipschitz constraint in this function. Initially, a radical weight clipping strategy, as suggested in \cite{Arjovsky-WGAN}, was applied. Later, we replace weight clipping with a more elaborated gradient penalty approach \cite{gulrajani2017improved}. 
It is worth noting that in our experiments, none of them produced a significant enhancement in the convergence of the GAN training and in many occasions, we observed that the gradient penalty heuristic even produced significant oscillations. Therefore, we finally chose a WGAN architecture with no additional strategy to enforce the Lipschitz constraint and the discriminator was optimized using only small learning rates and a new adaptive mini-batch procedure as heuristics to avoid reaching mode collapse situations. 

We selected fully connected neural networks (FCNNs) as the architectural model for both the discriminant and the generative networks. This decision was based on the observation that the statistical nature of the 4 features to be synthetically replicated did not exhibit any topological structure or time relationship among them and therefore, convolutional (CNNs) or recurrent networks (e.g. LSTMs) respectively would not provide any advantage with respect to FCNNs.
We observed in preliminary experiments that very deep networks with a large number of hidden layers or units did not generate significant improvements in performance and on the contrary, they enlarged convergence times and produced non-negligible oscillations in the convergence during the training process.
This effect could be explained by the fact that the cryptomining classification problem does not need very complex models to obtain a decent accuracy \cite{pastor2020detection}. Therefore, we selected a moderate number of hidden layers (between 3 and 5) for generator and discriminator networks. 

To provide the generator with more complex nonlinear capabilities to learn how to fool the discriminator,  we used hyperbolic tangent and Leaky-ReLU functions \cite{maas2013rectifier} as activation functions in the neurons of its hidden layers. In the case of the discriminator only LeakyReLUs were used.
Regarding that the discriminator does not play as a direct critic as in standard GANs but as a helper for estimating the Wasserstein distance between real and generated data distributions, the activation of its output layer is a linear function.
As the generator has to produce synthetic samples close to the real data, we considered two possibilities for the activation functions of its output layer: linear and domain-customized functions. A detailed explanation of the rationale and  trade-offs of using domain-customized activation functions instead of linear functions is presented in the next subsection \ref{sec-sub:custom-ouput}.

In addition, we selected a well-known heuristic for training WGANs from the literature \cite{gulrajani2017improved}  and designed some new ones to test if they offered any advantage in the convergence of the training process or in the performance of the synthetic data during the attack detection process.
In subsection \ref{subsec:heuristics} we detail the applied heuristics: 
(i) adaptive mini-batches on training, (ii) noise addition, (iii) multiple-embedding and (iv) complementary traffic addition.


\begin{figure*}[!t]
\centering
\begin{subfigure}[t]{.245\textwidth}
\includegraphics[width=1\linewidth]{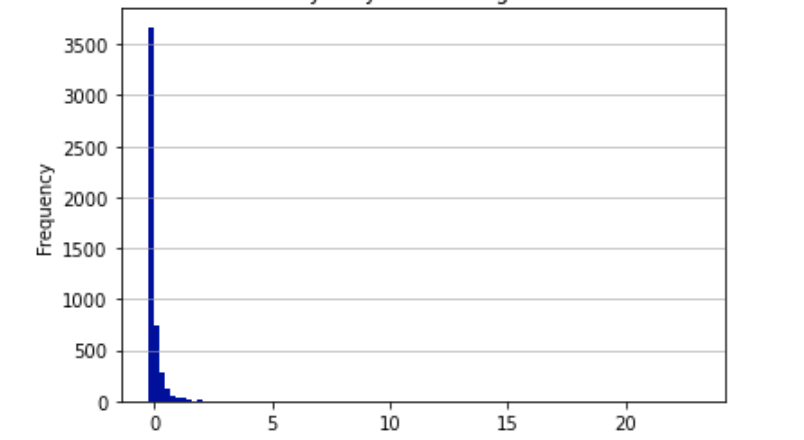} 
\caption{Label 0. Feature 1}\label{fig:distr-real-l0-f1}
\end{subfigure}
%
\begin{subfigure}[t]{.245\textwidth}
\includegraphics[width=1\linewidth]{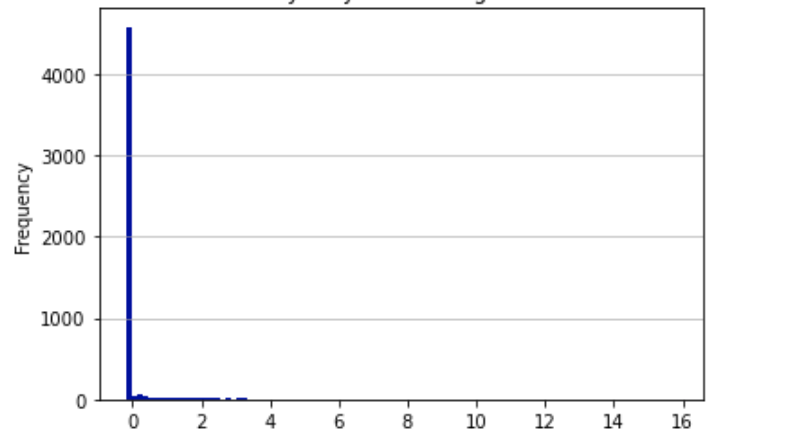} 
\caption{Label 0. Feature 2}\label{fig:distr-real-l0-21}
\end{subfigure}
%
%
\begin{subfigure}[t]{.245\textwidth}
\includegraphics[width=1\linewidth]{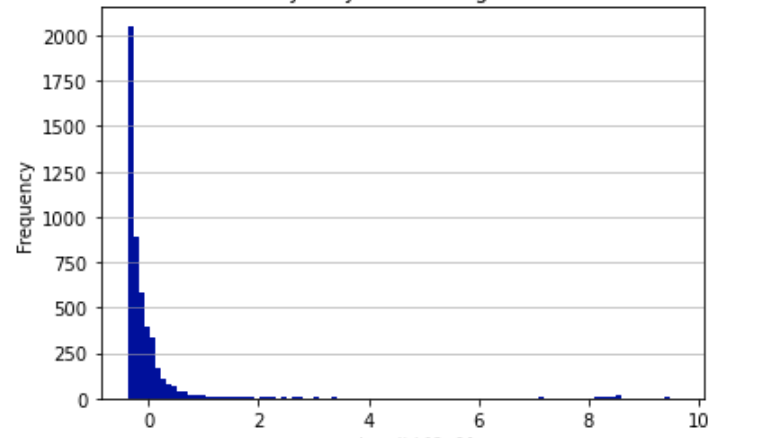} 
\caption{Label 0. Feature 3}\label{fig:distr-real-l0-f3}
\end{subfigure}
\begin{subfigure}[t]{.245\textwidth}
\includegraphics[width=1\linewidth]{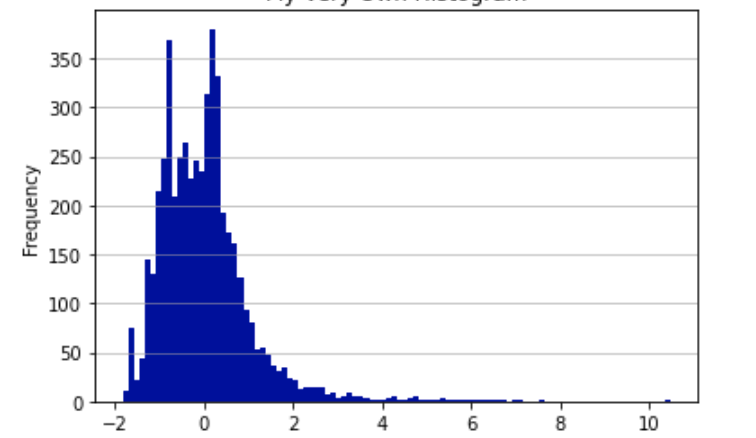} 
\caption{Label 0. Feature 4}\label{fig:distr-real-l0-f4}
\end{subfigure}

\medskip

\begin{subfigure}[t]{.245\textwidth}
\includegraphics[width=1\linewidth]{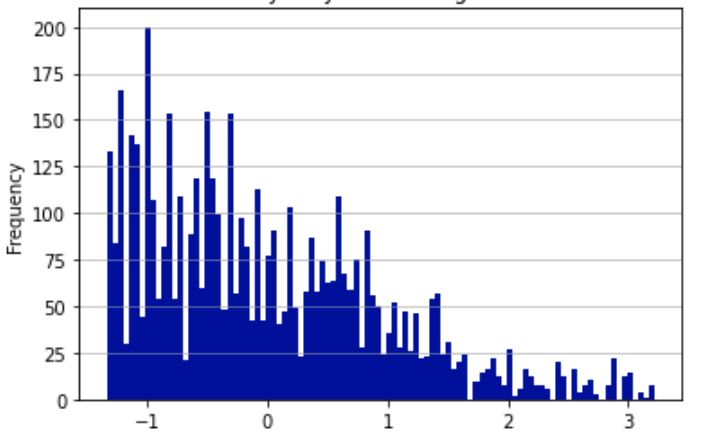} 
\caption{Label 1. Feature 1}\label{fig:distr-real-l1-f1}
\end{subfigure}
%
\begin{subfigure}[t]{.245\textwidth}
\includegraphics[width=1\linewidth]{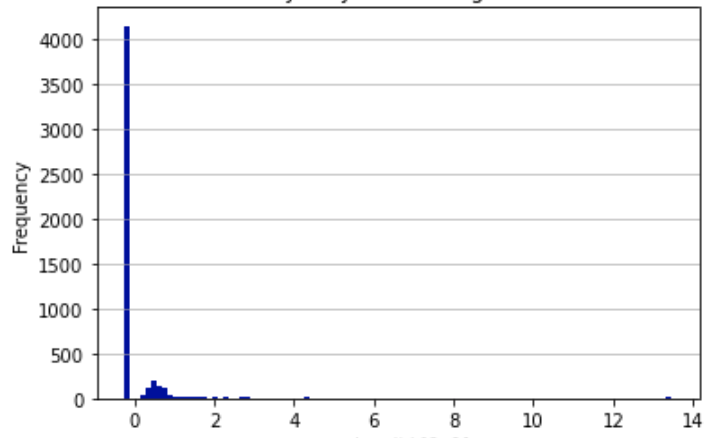} 
\caption{Label 1. Feature 2}\label{fig:distr-real-l1-f2}
\end{subfigure}
%
%
\begin{subfigure}[t]{.245\textwidth}
\includegraphics[width=1\linewidth]{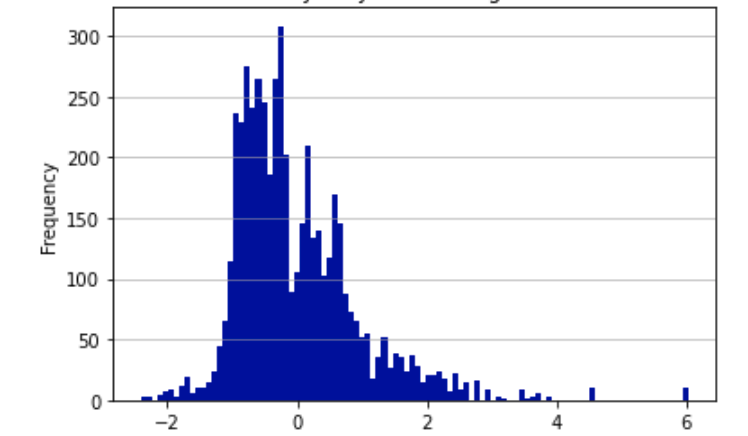} 
\caption{Label 1. Feature 3}\label{fig:distr-real-l1-f3}
\end{subfigure}
\begin{subfigure}[t]{.245\textwidth}
\includegraphics[width=1\linewidth]{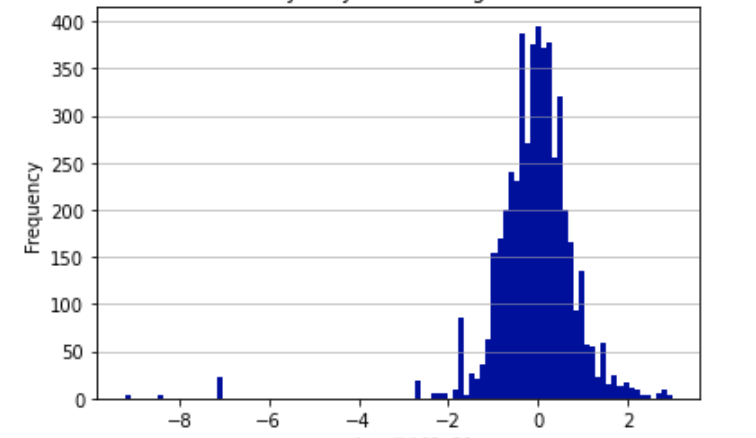} 
\caption{Label 1. Feature 4}\label{fig:distr-real-l1-f4}
\end{subfigure}

\caption{Frequency distribution histogram of the 4 variables extracted from normal (Label 0) and cryptomining (label 1) traffic. Label 0 (\ref{fig:distr-l0-1}, \ref{fig:distr-l0-5}, \ref{fig:distr-l0-200}, \ref{fig:distr-l0-800} and \ref{fig:distr-l0-1500}) and label 1 (\ref{fig:distr-l1-1}, \ref{fig:distr-l1-5}, \ref{fig:distr-l1-200}, \ref{fig:distr-l1-800} and \ref{fig:distr-l1-1500}) }
\label{fig:real-data-distribution}
\end{figure*}

\subsection{Custom activation function for the real data domain}
\label{sec-sub:custom-ouput}

An important issue in the generation of synthetic replicas of network traffic variables is that real data are sometimes not normally distributed. In general, telecom domain variables used in ML are usually statistical data representing the evolution of flow variables such as counters, accumulators, and ratios that always take positive values. In our real-time experiments, flows are monitored periodically from start to finish and therefore, the collected values for some of these variables are not normally distributed and tend to follow an exponential distribution with many occurrences of values near to $0$ and a long tail of large values appearing very rarely. Figure \ref{fig:real-data-distribution} shows the frequency distribution histograms of the four variables extracted from the normal traffic and cryptomining connections we used in our experiments.
Generator networks usually have a linear activation in the neurons of their output layer, which produces output variables following a normal distribution with mean that of the distribution of the real data.
When the real data follow a different distribution (e.g. exponential), using linear activation functions in the output layer of the generator can produce synthetic data outside the domain of the real data. 
For example, if we consider a variable representing an accumulator, only positive values are possible in the domain of real data. However, the generator will produce synthetic data with the same mean as the real data distribution (exponential) but following a normal distribution that contains negative data not existing in the real domain (elements of the leftmost part of the bell-shaped distribution).
This anomaly can be observed graphically by superimposing the exponential and normal density curves  on the mean value. A set of points appears on the leftmost part of the bell-shaped curve but not on the exponential curve. These points will take negative values and therefore, they do not exist in the real data domain. Furthermore, the bell-shaped curve is not containing the rare elements appearing in the rightmost part of the long tail of the real data. 
It is worth noting  that this problem has not attracted much attention in the literature since most of the work on GANs has been done for image generation where a normal distribution of pixel values is tolerated quite well by the human eye.

To mitigate this problem, we propose to use specific activation functions in the output neurons of the generator to adjust as much as possible the data distributions of the generator outputs to the statistical distribution of the real variables and thus, avoid the generation of negative values outside the domain of such variables.
To try to replicate variables that follow an exponential distribution, we propose to use ReLU functions in the generator as activation functions of the neurons at the output layer. The ReLU function only generates positive values due to its non-linear behaviour (the output is the input value for positive values and $0$ for negative values). In order to provide a smoother transition at values close to $0$, we experimentally observed that a Leaky-ReLU function with a very small slope for negative values performed better than a pure ReLU function. 
It is worth noting that the use of a Leaky-ReLU function will generate a marginal number of samples with negative values that can be easily filtered out later in a post-processing step. Nevertheless, further research work should investigate new activation functions to perfectly match the statistical distribution of real variables.

\subsection{Heuristics}
\label{subsec:heuristics}

We designed three novel mechanisms and applied a well-known heuristic based on adding noise to the discriminator and tested each of them to see if they could impact on the training convergence or the quality of the generated synthetic data. These heuristics are the following:
 (i) an adaptive number of mini-batch cycles are applied to the training of discriminator and generator networks to avoid the occurrence of the so-called collapse mode and balance the learning speed of both networks, (ii) different types of noise are added to the discriminator inputs to slow down its learning speed, (iii) a multi-point embedding of a single class is added to the input layer of the generator for augmenting the variety of the latent noise vector and (iv) real traffic of the class not modeled in the GAN is added jointly with the set of fake examples to slow down the learning process of the discriminator.

\subsubsection{Adaptive mini-batches on training}

This heuristic aims to avoid the occurrence of the so-called collapse mode during training when one network learns faster than the other, which finally produces that the slower network cannot learn any more. 
This anomaly was observed in preliminary experiments when a generator was not able to fool the discriminator with any synthetic example or on the contrary, when the discriminator could not identify any synthetic example as fake. 

As previously described, the standard GAN training consists of a mini-batch  stochastic gradient  descent  training  algorithm in which each mini-batch training consists of the execution of one train\_batch for the discriminator followed by one train\_batch for the generator. In this way, the discriminator  and generator networks are trained in such a  way  that  when  one  network  is  trained  the  learning  of  the other is blocked. 
We propose a modified training procedure  to avoid the blocking problems that we observed during preliminary experiments when we used such standard training process. 
The modified training procedure is described in Figure \ref{algo} and consists on  executing a variable number of train\_batch for the discriminator and generator networks. Each network is trained until a minimum value of successful elements are correctly classified at the end of the mini-batch training. For the discriminator, we force extra train\_batch cycles until the ratios of real and fake examples that are correctly classified are greater than two pre-established thresholds. For the generator, the ratio of incorrectly classified samples (i.e., fake samples that are considered as real samples by the discriminator) must be greater than a predetermined threshold at the end of the mini\_batch training. Otherwise, additional training cycles are added. 
This  heuristic avoids situations in which the discriminator or the generator learns faster than its opponent and finally blocks the learning evolution of its counterpart. In this situation, some of these ratios reach zero and the slow-learning network is no longer able to learn anymore during subsequent training steps and therefore, improve these ratios.

During our experiments, we observed that using moderate ratios of 0.1 or greater for the generator produced oscillations in the convergence of the training process that disappeared when smaller values were used (0.05 for the generator and 0.1 for both ratios of the discriminator).
Although no significant improvement on performance or convergence speed was observed in our experiments when adaptive mini-batches were activated and these small ratios were used, the previously observed blocking situations disappeared.

\begin{algorithm}

    \captionof{figure}[Enhanced GAN training procedure with adaptive mini-batches]{Enhanced GAN training procedure with adaptive mini-batches\label{algo}}
    \footnotesize
    \begin{algorithmic}[1]
 
            \Procedure{GAN\_train\_batch}{}
            
            \Repeat
                \State $train\_batch (discriminator)$
                \State $preds \gets  discriminator.predict(sample(real,synthetic)) $
            \Until {$TP(preds)$ and $TN(preds)$ are in range}
            
            \Repeat
                \State $train\_batch (generator)$
                \State $preds \gets discriminator.predict(generator.predict(noise))$
                \State $ratio\_fake\_pass \gets len(preds = real\_label)/len(preds)$
            \Until {$ratio\_fake\_pass > min\_ratio\_fake\_pass$}
            
            \EndProcedure
            
    \end{algorithmic}
    
\end{algorithm}

\subsubsection{Noise addition}

 This heuristic aims to slow down the learning rate of the discriminator in order that the generator can learn how to fool the discriminator in each mini-batch training cycle. It is worth noting that this heuristic works in a complementary way to the adaptive learning rates that the optimization algorithm applies to each mini-batch training.

In order to avoid disjoint distributions, the authors in \cite{gulrajani2017improved} suggested to add continuous noises to the inputs of the discriminator to artificially "spread out" the distribution and to create higher chances for two probability distributions to have overlaps. To this end, we add several types of noises that are configured as hyperparameters during the training. Three different types of noise (Uniform or Gaussian with mean $0$ and configurable standard deviation) can be stacked: noise added to (i) fake examples, (ii) to all  (real and fake) examples; and (iii) a configurable percentage of fake and real labels are changed to its opposite value.

Our preliminary experiments revealed that even moderate amounts of noise did not allow the generator to learn adequately and the quality of the obtained synthetic data was really poor as reflected by the distance metrics with respect to the real data. Moreover, when the synthetic data was used for substituting real data in the training of a traffic classifier, the $F_1$-score obtained was also smaller. When a large amount of noise is added, the rightmost values of the long tail distribution of real variables tend to disappear in the discriminator and therefore, they will not be learnt by the generator. In fact, it can be observed that the synthetic distributions of these variables tended to be grouped around the mean of the variables. 
On the contrary, small amounts of noise did not produce any bad effect on the convergence or quality of synthetic data, but neither did they generate any significant improvement. 

\subsubsection{Multi-point single-class embedding}
We designed a new way to input latent noise to the generator. Instead of generating a noise vector from a uniform distribution in a large interval of values $\interval{-K}{K}$, we provided latent vectors generated uniformly at random from smaller intervals of values and centered at different points in the latent space that also were selected uniformly at random. 
An additional hyperparameter allows to also train the optimal location of the centroids in the latent space. The rationale of this heuristic was to explore whether it was easier to train a GAN with small random bubbles of latent vectors than to use a single latent vector from a larger range of random values.

We implemented the centroids of these random bubbles using an embedding layer of the same dimension than the latent vector. The input to the embedding layer was a number representing the bubble and the output were the coordinates of this bubble in the latent space. The latent noise is generated by drawing a sample from a normal or uniform distribution. This value is added to the bubble centroid coordinates to generate a point (latent vector) around this location to be fed into the generator network.
The embedding layer weights can also be learnt during training to find an optimal location of the bubbles in the latent space, or they can be chosen at random and frozen during GAN training.

Although this heuristic was only superficially investigated  and we did not observe any significant improvement in convergence or synthetic data quality, future work should explore more carefully the implications of using these latent bubbles.

\subsubsection{Complementary data}
When training the discriminator using network traffic of type X, it is possible to mix a configurable ratio of real examples of the other type of traffic Y  with fake examples obtained from the generator. The rationale of this heuristic was to avoid that the discriminator overfitted on the fake data produced by the generator at each mini-batch as it has to also learn to differentiate the other type of real traffic Y. During preliminary experiments, we did not observe any significant advantage when this heuristic was included with different ratios of Y examples ranging from $0.1$ to $0.5$.

\section{Performance metrics}
\label{sec:metrics}

We propose to evaluate GANs performance using two different types of metrics. The first set of metrics is inspired by the $L^1$ functional distance and the Jaccard coefficient and aim to quantify the similarity of the synthetic data with respect to the real data from a statistical perspective and considering the joint distribution of data features. On the other hand, the second set of metrics attempts to quantify the performance of synthetic data when it is used as a substitute for real data in the training of a ML classifier that is trained to distinguish between normal and cryptomining traffic. 
These two types of metrics will be used to compare the similarity between real and synthetic distributions and  will also be applied to implement a stopping criterion for GAN training to select  generators that produce high-quality synthetic data.

To the best of our knowledge, this is the first time that $L^1$ metric and Jaccard coefficient are used for defining metrics to compare synthetic and real data in GANs and furthermore, there is no other work that proposes to use the two types of metrics to implement a stopping criterion for GAN training.

\subsection{$L^1$-metric and Jaccard index}

The first two metrics we introduce try to measure the difference between the probabilistic distribution of the real data and the one of the synthetic data. They are based on two well-known statistical coefficients applied for hypothesis testing and probabilistic distances.

For the convenience of the reader, we briefly review some relevant definitions. Suppose that $X$ and $Y$ are two independent continuous $d$-dimensional random vectors with probability density functions $f_X, f_Y: \mathbb{R}^d \to \mathbb{R}$. To measure the distance between $X$ and $Y$, we can consider the $L^1$-metric between their density functions as
$$
    d_{L^1}(X,Y) = \int_{\mathbb{R}^d} |f_X(s) - f_Y(s)|\,ds.
$$
Notice that $d_{L^1}(X,Y) = 0$ if and only if $f_X = f_Y$ almost sure and thus $X = Y$ almost sure.

Additionally, we can also compare the supports of $X$ and $Y$ through the standard Jaccard coefficient \cite{tanimoto1958elementary}. Let $\textrm{supp}(f_X)$ be the support of the function $f_X$, that is, the closure of the set of points $s \in \mathbb{R}^d$ such that $f_X(s) \neq 0$. Then, the Jaccard index of $X$ and $Y$ is given by
$$
    J(X,Y) = \frac{|\textrm{supp}(f_X) \cap \textrm{supp}(f_Y)|}{|\textrm{supp}(f_X) \cup \textrm{supp}(f_Y)|},
$$
where $|A|$ denotes the Lebesgue measure of a measurable set $A \subseteq \mathbb{R}^d$. Notice that $\textrm{supp}(f_X) \cap \textrm{supp}(f_Y) \subseteq \textrm{supp}(f_X) \cup \textrm{supp}(f_Y)$ so $0 \leq J(X,Y) \leq 1$ and, the larger the index, the more similar the supports. Indeed, perfect agreement of the supports is achieved if and only if $J(X,Y) = 1$.

Nevertheless, in this form these ideas can only be applied theoretically in a scenario where the density functions are perfectly known. This obviously does not hold in a practical situation. However, the previous definitions can be straightforwardly extended to the sampling setting by replacing the probability density function by the histogram of a sample.

Suppose that we have samples $x_1, \ldots, x_n$ of a random vector $X$, with $x_i = (x_i^1, \ldots, x_i^d)$. From them, we can estimate the density function of $X$ through the histogram function $h_X: \mathbb{R}^d \to \mathbb{R}$. For this purpose, choose a partition of $\textrm{supp}(f_X) \cup \textrm{supp}(f_Y)$ into $d$-dimensional cubes
$$  
    \textrm{supp}(f_X) \cup \textrm{supp}(f_Y) = \bigsqcup_{k=1}^\ell C_k.
$$
A common choice for this partition is constructed as follows. Let $m^j = \min_i(x_i^j)$ and $M^j = \max_i(x_i^j)$ be the maximum and minimum of the estimated support of the $j$-th component of $X$. Take an uniform partition $m^j = s_0^j < s_1^j < \ldots < s_w^j = M^j$ of the interval $[m^j,M^j]$. Then, the cubes of the partition are given by the product of intervals $C_{i_1, i_2, \ldots, i_d} = [s_{i_1-1}^1, s_{i_1}^1) \times [s_{i_2-1}^2, s_{i_2}^2) \times \ldots \times [s_{i_d-1}^d, s_{i_d}^d)$ for $1< i_1, \ldots i_d \leq w$.

In any case, given a partition $C_k$, we define the histogram function to be
\begin{align}
\label{eq:1}
     h_X(s) = \frac{1}{n}\sum_{k = 1}^{\ell} \left(\sum_{i=1}^n\chi_{C_k}(x_i)\right)\chi_{C_k}(s),
\end{align}
where $\chi_{C_k}: \mathbb{R} \to \mathbb{R}$ is the characteristic function of the cube $C_k$, that is, $\chi_{C_k}(s) = 1$ if $s \in C_k$ and is $0$ otherwise. In other words, if $s$ belongs to the bin $C_k$, then $h_X(s)$ is the average of the number of samples $x_j$ lying in the $d$-dimensional cube $C_k$. Recall that the integral of $h_X$ is strongly related to the empirical cumulative probability function which, by the Glivenko-Cantelli theorem \cite{van1996glivenko}, converges almost surely to the real cumulative probability function. In this way, for large samples, it may be expected that $h_X$ estimates rather faithfully the real density function $f_X$.

In particular, this histogram function allows us to estimate the aforementioned metrics. Suppose that we have samples $x_1, \ldots, x_n$ and $y_1, \ldots, y_m$ of random variables $X$ and $Y$, respectively. Choose a common partition $\{C_k\}_{k=1}^{\ell}$ of the union of the supports of the samples. Then, we define the sampling $L^1$-metric to be
$$
    d_{L^1}^{smp}(X,Y) =  \int_{\mathbb{R}^d} |h_X(s) - h_Y(s)|\,ds =  \sum_{k = 1}^{\ell} |h_X(C_k)-h_Y(C_k)|\textrm{Vol}(C_k) =  L\sum_{k = 1}^{\ell} |h_X(C_k)-h_Y(C_k)|.
$$
Here, $\textrm{Vol}(C_k)$ denotes the Lebesgue measure of the cube (its volume), $h_X(C_k)$ refers to the value of $h_X$ at any point of the cube $C_k$ (recall that $h_X$ is constant on the cubes). Finally, in the last equality, we have supposed that the partition is uniform and we set $L = \prod_{j=1}^d(M^j-m^j)/w$. In analogy with the the purely probabilistic case $d_{L^1}^{smp}(X,Y) = 0$ if and only if the number of samples of $X$ and $Y$ in each are equal, if the bins of the partition are the same.

In a similar vein, the Jaccard index can be estimated from the histograms. Let $\textrm{supp}(h_X), \textrm{supp}(f_Y)$ be the supports of the histograms. Then we define the sampling Jaccard index as
$$
    J^{smp}(X,Y) = \frac{|\textrm{supp}(h_X) \cap \textrm{supp}(h_Y)|}{|\textrm{supp}(h_X) \cup \textrm{supp}(h_Y)|}.
$$
Again, this coefficient takes values in the interval $[0,1]$ and the larger the value of $J(X,Y)$ the more similar the empirical supports.

In our particular case of GANs, we shall apply these coefficients to measure the similarity between the real and the synthesized data. Let $x_1, \ldots, x_n$ the real instances of the dataset $X$ to be replicated. Given a generator network $G$, we extract a sufficiently large sample $y_1^G, \ldots, y_m^G$ of generated data $Y$. Then, the $L^1$ metric and the Jaccard index of the generator $G$ are just
$$
    d_{L^1}(G) = d_{L^1}^{smp}(X,Y^G), \qquad J(G) =  J^{smp}(X,Y^G).
$$

\subsection{Nested ML performance}\label{sec:nested-ml}

The second set of metrics attempts to quantify the performance of synthetic data when it is used as a substitute for real data for training a ML classifier to distinguish between normal and cryptomining internet traffic.

To be precise, let $C: \mathbb{R}^d \to \left\{0,1\right\}$ be a binary classifier. It attempts to take an instance $x = (x^1, \ldots, x^d) \in \mathbb{R}^d$ (which, in our case, represents the $d$ features of a internet connection) and to predict its class $C(x) \in \left\{0,1\right\}$ (the type of traffic in our setting). Once the classifier $C$ has been trained, its accuracy can be measured against the test split of the dataset, where the real classes $\mathtt{Y}^{\textrm{test}}=(y_1, \ldots, y_n)$, with $y_i \in \left\{0,1\right\}$, of a bunch of instances $\mathtt{X}^{\textrm{test}} = (x_1, \ldots, x_n)$, with $x_i \in \mathbb{R}^d$, are known. In that case, we define precision and recall as the quantities
$$
    \textrm{Precision}(C) = \frac{\left|\left\{x_i \in \mathtt{X}^{\textrm{test}}\,|\, C(x_i) = 1 \textrm{ and } {y}_i = 1\right\}\right|}{\left|\left\{x_i \in \mathtt{X}^{\textrm{test}}\,|\,C(x_i) = 1\right\}\right|}, \qquad 
	\textrm{Recall}(C)=\frac{\left|\left\{x_i \in \mathtt{X}^{\textrm{test}}\,|\, C(x_i) = 1 \textrm{ and } {y}_i = 1\right\}\right|}{\left|\left\{x_i \in \mathtt{X}^{\textrm{test}}\,|\,{y}_i = 1\right\}\right|} .
$$

Here, $|X|$ stands for the number of elements of the set $X$. In other words, $1-\textrm{Precision}(C)$ is the rate of false positives and $1-\textrm{Recall}(C)$ is the rate of false negatives of the class $\lambda$.  In general, to combine both coefficients, it is customary to consider the $F_1$ as the harmonic mean
$$
	F_1\textrm{-score}(C) = 2 \frac{\textrm{Precision}(C)\cdot \textrm{Recall}(C)}{\textrm{Precision}(C)+\textrm{Recall}(C)}.
$$

Additionally, these metrics can be complemented with the so-called confusion matrix. It is a $2 \times 2$ matrix that compares the real labels of each instance with the predicted label, in the form
\small
$$
    \begin{pmatrix}
    \left|\left\{x_i\,|\, C(x_i) = 0 \textrm{ and } {y}_i = 0\right\}\right| & \left|\left\{x_i\,|\, C(x_i) = 1 \textrm{ and } {y}_i = 0\right\}\right| \\
    \left|\left\{x_i\,|\, C(x_i) = 0 \textrm{ and } {y}_i = 1\right\}\right| & \left|\left\{x_i\,|\, C(x_i) = 1 \textrm{ and } {y}_i = 1\right\}\right|
    \end{pmatrix}
$$
\normalsize
In other words, the diagonal entries are the correct classified instances and the off-diagonal entries are the false positives (upper-right corner) and the false negatives (botton-left corner). In this way, the confusion matrix allows us to identify more precisely the flaws of the classifier $C$, in comparison with the precision, recall and $F_1$ measures, which are raw means.

With this notions at hand, the quality of a GAN will be evaluated as follows. Suppose that, as explained in Section \ref{subsec:architecture}, we have trained GANs $(\Lambda_0, G_{0}, D_0)$ and $(\Lambda_1, G_1, D_1)$ to synthesize data with label $0$ (real traffic) and $1$ (cryptomining traffic) respectively. Choose $N, M > 0$ and draw samples $x_1^0, \ldots, x_N^0$ and $x_1^1, \ldots, x_M^1$ of the latent spaces $\Lambda_0$ and $\Lambda_1$ respectively. Then, using the generators $G_0$ and $G_1$, we create a new fully synthetic training dataset
$$
\mathtt{X}^{\textrm{train}} = \left\{G_0(x_1^0), \ldots, G_0(x_N^0), G_1(x_1^1), \ldots, G_1(x_M^1)\right\}, \qquad
\mathtt{Y}^{\textrm{train}} = \{\underbrace{0, \ldots, 0}_{N \textrm{ times}}, \underbrace{1, \ldots, 1}_{M \textrm{ times}}\},
$$
with $N + M$ instances.

With this new dataset $(\mathtt{X}^{\textrm{train}}, \mathtt{Y}^{\textrm{train}})$, we train a standard ML classifier $C$ (say, a random forest classifier). Then, screening the precision, recall, and $F_1$-score of $C$ against a test split made of real data, we are able to measure the quality of the generated data: the higher these measures, the better the data. Hence, large values of these coefficients point out that the synthetic data generated by $G_0$ and $G_1$ can be used to faithfully substitute the real instances. Observe that no real traffic is used for such training purposes, although real traffic is always used for testing. 
As previously stated, this is a differentiating characteristic of our work with respect to existing solutions. Data augmentation solutions are proposed in these previous works, where synthetic data is mixed with real data during training, generating data privacy breaches as real data is used.

Several variants of this proposal can be considered. First, instead of creating the dataset with fully-trained GANs $(\Lambda_0, G_{0}, D_0)$ and $(\Lambda_1, G_1, D_1)$, we can compute these coefficients at each of the training epochs of the GAN. 
In this way, we are able to screen the evolution of the training and to relate it to the quality of the generated data. In particular, this idea enables a stopping criterion: when the GAN training epochs do not produce any significant enhancement, the training process is stopped.
As previously commented, one of the current open issues of GANS is how to optimize their training as oscillatory behaviours appear frequently during training. Moreover, none of the related works listed in Section \ref{sec:related_work} explain what stopping criterion they applied to obtain their fully trained GANs. Therefore, our proposal for measuring GAN performance in ML tasks at each epoch provides a way to implement such stopping criterion.
 
Additionally, we can also evaluate the marginal quality of each of the generators. In the previous approach, we generated the dataset $(\mathtt{X}^{\textrm{train}}, \mathtt{Y}^{\textrm{train}})$ by using synthetic samples both for label $0$ and $1$. 
However, if we want to test the quality when generating only one of the labels, say label $0$, the dataset $(\mathtt{X}^{\textrm{train}}, \mathtt{Y}^{\textrm{train}})$ can be also created by mixing synthetic samples of label $0$ with real samples of label $1$. 
In this way, the corresponding ML accuracy coefficients will only measure the ability of $G_0$ to generate label $0$, regardless of the fitness of $G_1$.

Regarding that we will use two different WGANs for each type of traffic, the first approach would imply to evaluate the ML performance of $G_0$ at each epoch with all generators obtained during $G_1$ training and conversely, at each $G_1$ epoch we should combine its generator with all $G_0$ generators to measure ML performance. Hence, assuming we trained a pair of GAN for $j$ and $k$ epochs respectively, we would require $j \times k$ evaluations of the ML task (training and testing) to obtain the full set of metrics. 

In order to implement the stopping criterion more efficiently, we opted for the second approach, in which we evaluate separately the marginal quality of each of the generators. 
In this way, each WGAN can be trained in parallel without requiring the other to evaluate their joint performance and therefore, each training can be stopped at different epochs when no significant enhancement is observed. 
When both WGANs are trained and the joint performance has to be computed, instead of generating the Cartesian product ($j \times k$) of the two sets of generators and running the corresponding ML evaluations, we  observed experimentally that drawing roughly a dozen samples by choosing uniformly at random one generator of each type of traffic tends to produce  results equivalent to the brute force approach of trying all possible combinations. 
It is worth noting that more elaborated strategies can be applied as ordering the generators of each type of traffic by some metric (e.g., $F_1$-score) and choosing generators at random only from the subset containing the best generators.

Finally, notice that there is plenty of freedom for choosing the number of generated samples $N$ and $M$. A first decision would be to choose $N$ and $M$ to be in the same range as the number of samples in the original dataset. This leads to a synthetic dataset with very similar characteristics to the original one in terms of balancing between classes. However, other set-ups can be tested like drastically increasing the number of samples in the generated dataset or to balance the number of instances of each class to ease the task of the classifier. These possibilities will be explored in Section \ref{sec:empirical}. It is worthy to mention that, even though the potential balance between classes achieved with this method is similar to what can be obtained with data augmentation procedures, the proposed solution is way stronger than standard data augmentation: the generated data is not a simple enrichment of the original dataset but a completely new dataset.
    
\section{Empirical evaluation}
\label{sec:empirical}

We designed a set of experiments to demonstrate that GANs can be utilised for generating high-quality synthetic data that replicates the statistical properties of real data while maintaining their privacy.
Furthermore, in these experiments we aim to prove that synthetically generated data can be utilised for totally substituting real data in ML training processes while keeping the same performance than ML models trained with real.

In this section, we first summarize the testbed on which we conducted our experiments and how we collected the data, then the experimental setup is detailed and finally, the experimental results are depicted. 

\subsection{Testbed for data collection}
\label{subsec:testbed}
The data sets used in our GAN experiments were previously generated in a realistic network scenario called the Mouseworld lab \cite{pastor2018mouseworld}. The Mouseworld is a network digital twin created at Telefónica R+D facilities that allows deploying complex network scenarios in a controlled way. In the Mouseworld, realistic labeled data sets can be generated to train supervised ML components and validate both supervised and unsupervised ML solutions. The Mouseworld Lab provides a way to launch real clients and servers, collect the traffic generated by them and the recipients outside the Mouseworld on the Internet, and add labels to the traffic automatically. 

To obtain the training and testing datasets used in our GAN experiments, we deployed in the Mouseworld thirty virtual machines for the generation of regular traffic (i.e.\ web, video and shared-folder flows) to internal Mouseworld servers and to external servers located in the Internet. The IXIA BreakingPoint tool was also configured to generate and inject synthetic patterns of various Internet network services (web, multimedia, shared-folder, email and P2P). All traffic generated was composed of encrypted and non-encrypted flows. 
In addition, we created three cryptomining Linux virtual machines in which we installed well-known cryptomining clients for mining the Monero cryptocurrency, which is commonly used for illegal purposes. The cryptomining clients were connected to public mining pools using non-encrypted TCP and encrypted TLS connections. 

We deployed in the Mouseworld four experiments with different cryptomining protocols \cite{pastor2020detection}. Each experiment was run for one hour with an average packet rate of approximately 1000 packets per second, which generated data sets with 8 millions of flow-based entries containing statistics of the TCP connections of which 4 thousands were related to cryptomining connections. Normal connections were labelled with $0$ and cryptomining ones with $1$. The four obtained data sets were split in two separate subsets for training and testing purposes. Specifically, the data sets from experiments 1 and 4 were joined in DS1 (training) data set and the other two data sets collected in experiments 2 and 3 were combined into DS2 (testing) data set. In this way, DS1 and DS2 can be considered of the same nature as they contain similar percentages of encrypted and non-encrypted traffic, types of internet services and cryptomining protocol flows. 
Considering the small amount of traffic generated by cryptomining protocols compared to normal traffic, it is worth noting that an imbalance in the number of cryptomining flows versus normal traffic appeared in both data sets. 
In each experiment, around $400K$ samples of label $0$ (normal traffic) appeared against  $4K$ instances of label $1$ (cryptomining traffic).

As previously commented, the goal of this work is to generate synthetic traffic that can substitute real traffic for training a ML-based traffic classifier that predicts with similar performance than a model trained with real data.
Nowadays, most machine and deep learning techniques use flow descriptions as input to machine learning models. These descriptions are composed of a set of features that are typically statistical data obtained from externally observable traffic attributes such as duration and volume per flow, inter-packet arrival time, packet size and byte profiles. Therefore, our GAN experiments will try to generate synthetic replicas of some statistical data that can be used as features to be input to a machine learning based network traffic classifier. In the data gathering experiments, a set of 59 statistical features were extracted from each TCP connection.

We selected a reduced set of 4 of these 59 features for our GAN experiments (the rationale of this choice is detailed in Section \ref{sec:problem-setting}): (a) number of bytes sent from the client, (b) average round-trip time observed from the server, (c) outbound bytes per packet, and (d)  ratio of packets\_inbound / packets\_outbound.

Therefore, we obtained a reduced version of DS1 and DS2 containing only the 4 selected features for training and testing our GANs. 
It is worth noting that other subsets of features were considered in preliminary experiments obtaining GANs with a similar performance to the ones shown in this paper.

\subsection{Experimental setup}
\label{subsec:experimental-setup}
To perform our experiments, we designed and trained independently two WGANs, one for each type of traffic, applying the set of hyperparameters detailed in Table \ref{table:hyperparameters}. Using this table as a reference, and taking into account that in general the training process of a single WGAN took one week on average, we performed a blind random search in the hyperparameter space but guided by the $F_1$-score obtained in a nested ML-model that was executed after each train epoch and evaluating the marginal quality of the generator at each epoch (see subsection \ref{sec:nested-ml}). 
In this way, we were able to independently adjust most of the parameters and observe whether or not the introduced modifications in a hyperparameter generated an improvement in the $F_1$-score of the classifier. The results are consistent throughout different executions, and the top generative models in different runs of the random search algorithm return similar performance metrics.

As a nested ML classifier, we used a Random Forest model with $300$ trees, which proved to be the best solution for classification when trained with the original (non-synthetic) dataset (c.f.\ Section \ref{sec:experiments-real} ). As shown in a previous work that used the real datasets DS1 and DS2 for training and testing\cite{pastor2020detection}, the performance of neural networks-based classifiers was quite poor as these models showed significant overfitting even after applying regularization procedures. On the contrary, Random Forest models exhibited very good performance. 

For each label, the WGAN selected was the one that obtained the best $F_1$-score for the nested classifier in any of its epochs.
It is worth noting that this method allows us to find the set of hyperparameter values that produces the best performance on a single WGAN for a type of traffic and therefore, it is not guaranteed that the best WGAN for label ``0'' (normal traffic) works well in combination with the best WGAN for label ``1'' (cryptomining traffic). Moreover, we observed that combining the generators of each WGAN that obtained the best $F_1$-score during partial nested evaluations, did not guarantee to obtain the best synthetic dataset when used in a nested evaluation of both types of traffic. In fact, we ran a simple heuristic to select pairs of generators that produced decent results during the nested evaluation without testing all possible combinations of generators from the two WGANS.  
These limitations highlight that future works should explore methods that enable the search of the best hyperparameters in both WGANs at the same time.

\begin{table*}[!t]
\centering
\caption{GAN hyperparameters.}
\label{table:hyperparameters}
\resizebox{0.67\linewidth}{!}{%
\begin{tabular}{>{\hspace{0pt}}m{0.0027\linewidth}>{\hspace{0pt}}m{0.184\linewidth}|>{\hspace{0pt}}m{0.272\linewidth}|>{\hspace{0pt}}m{0.184\linewidth}|}
 & \multicolumn{1}{>{\hspace{0pt}}m{0.204\linewidth}}{} &  & \textbf{Range of values} \\ 
\cline{2-4}
 & \textbf{Generator/discriminator}\par{}\textbf{FCNN architectures} & \textit{\# layers} & {[}2..6] \\ 
\cline{2-4}
 &  & \textit{\# units per layer} & {[}100..10000] \\ 
\cline{2-4}
 & \textbf{Generator} & \textit{Output Activation~} & (linear,custom) \\ 
\cline{2-4}
 &  & Output filtering with discriminator \par{}(\textgreater{}0, percentile) & (True,False)\par{}{[}0..100] \\ 
\cline{3-4}
 &  & \textit{Latent vector }\par{}\textit{with Embedding (categories)} & (True,False)\par{}{[}1..20] \\ 
\cline{3-4}
 &  & latent vector & Fixed 123 \\ 
\cline{3-4}
 &  & noise for latent vector (distr,std) & (normal, uniform)\par{}std=[0.1..100] \\ 
\cline{3-4}
 &  & batch normalization & {[}True..False] \\ 
\cline{3-4}
 &  & regularization: L2, dropout & Fixed values (0,0) \\ 
\cline{3-4}
 &  & learning rate & Default value (0.001) \\ 
\cline{3-4}
 &  & LeakyRelu alpha & Fixed value (0.15) \\ 
\cline{3-4}
 &  & \textit{Percentage of tanh/LeakyRelu }\par{}\textit{in internal units} & {[}0..100] \\ 
\cline{2-4}
 & \textbf{Discriminator}\par{} & \textit{Noise in fakes }\par{}\textit{(distribution,std)} & (normal,uniform)\par{}{[}0..20] \\ 
\cline{2-4}
 &  & \textit{Noise in all examples}\par{}\textit{(distribution,std)} & (normal,uniform)\par{}{[}0..20] \\ 
\cline{3-4}
 &  & \textit{Ratio Label change} & {[}0..20] \\ 
\cline{3-4}
 &  & batch normalization & {[}True..False] \\ 
\cline{3-4}
 &  & regularization: L2, dropout & {[}0..2], [0..30] \\ 
\cline{3-4}
 &  & LeakyRelu alpha & Fixed value (0.2) \\ 
\cline{3-4}
 &  & learning rate & {[}0.0001..0.001] \\ 
\cline{2-4}
 & \textbf{Adaptive~mini-batch} & \textit{generator. ratio fake pass} & Fixed (0.3) \\ 
\cline{2-4}
 &  & \textit{discriminator. ratio TP} & Fixed (0.01) \\ 
\cline{3-4}
 &  & \textit{discriminator. ratio TN} & Fixed (0.01) \\
\cline{3-4}
\end{tabular}
}
\end{table*}
Hyperparameters in Table \ref{table:hyperparameters} are grouped in four categories: (1) common parameters of the FCNN architecture for generator and discriminator, (2) generator parameters, (3) discriminator parameters and (4) adaptive mini-batch training parameters.
The hyperparameter space is detailed in "Range of values" column. The text "Fixed value" indicates that the hyperparameter was explored in a preliminary phase before executing the random search and therefore, the value was  previously determined and no random search was performed on it. Ranges described by a list of values in brackets indicate that the random search was performed by randomly choosing an element from the list. Conversely, two values in square brackets denote the minimum and maximum of the range of values to be considered in the random selection.

 As optimization algorithms, we used Adam for generators and RMSProp for discriminators. The typical binary cross-entropy loss function was substituted by the Wasserstein loss. Assuming that the discriminator generates a positive value if an example is classified as a real example or negative if the example is considered as fake, the Wasserstein loss multiplies the output of the critic (i.e., the discriminator) by $-1$ (real examples) or $1$ (fake examples). 
 For label "0" (normal traffic), we set the size of each minibatch as a ratio ($0.002$) of the total number of examples ($400,000$) and for label "1" (cryptomining traffic), we set this ratio to $0.02$ of the total number of samples ($4,000$).  
 
 Although it has been reported that activating batch normalization in generators can produce correlations in the generated samples, we decided to include it as a hyperparameter after observing that generators without batch normalization exhibited a lack of convergence on many occasions. 

To provide the generator with more complex non-linear capabilities to learn how to cheat the discriminator, we used Leaky-ReLU activation functions  with their slope parameter set to $\alpha = 1.5$ and hyperbolic tangent as activation functions in the neurons of its hidden layers. The ratio of hyperbolic tangents over the total of activation functions was a configurable hyperparameter. For the discriminator, we only used Leaky-ReLUs in the hidden layers setting a more aggressive value $\alpha = 0.2$.

\subsection{Experimental results}

In this section, we review the results obtained in the conducted experiments when real data sets are replaced by fully synthetic datasets. For this purpose, we shall compare the performance obtained by a ML classifier when trained with (1) real data (DS1 dataset), (2) data generated through a simple mean-based generator (working as baseline) and (3) a synthetic dataset generated with a standard WGAN (with no variants or improvements implemented). The effects of using an improved WGAN are analysed in section \ref{sec:improvements}.

\subsubsection{Real data} \label{sec:experiments-real}
A Random Forest was trained to classify the original real dataset into normal and cryptomining-based traffic. We chose Random Forest due to its well-known good performance in classification tasks and in particular, when cryptomining and normal traffic has to be classified\cite{pastor2020detection}. 
A hyperparameter tuning was conducted through a grid search on the number of classification trees used, ranging from 10 to 600 estimators. No depth limit was applied to trees.
This performance was evaluated against a validation split excerpted from the training data set. The experiments showed that using more than $300$ estimators did not produced any significant increase in $F_1$-score.
After this hyperparameter tuning, the model was re-trained with the whole training dataset (DS1) and evaluated against the test dataset (DS2). In order to analyse the impact of the decision threshold of the classifier on the number of false positives and negatives, several thresholds for the model to distinguish between the $0$ and the $1$ class were tested, with possible values $0.2, 0.4, 0.5$ (default value), $0.6$ and $0.8$. 

The results obtained in testing are shown in Table \ref{table:real}. 
and point out that the performance of the classifier against the original dataset is very high, with a $F_1$-score of $0.962$ with the best threshold (and slightly worse with the default threshold). 
Note that the confusion matrix shows that most of the wrongly classified instances are false negatives, i.e.\ cryptomining traffic (class $1$) samples that are classified as normal traffic (class $0$). Only a few false positives were observed. 
It should be noted that in certain scenarios this percentage of false positives may not be desirable as it would mean that users may be suffering from surreptitious use of their resources that would not be detected.
On the contrary, a non-negligible number of false negatives could imply extra efforts as false alarms will be raised in the detection system, which could imply  a individual treatment of each of them.

\begin{table*}[t!]
\parbox{.48\linewidth}{
\caption{Baseline results using real data for training.}
\label{table:real}
\centering
\resizebox{.99\linewidth}{!}{%
\begin{tabular}{|c|c|c|c|}
\hline \textbf{Dataset} & \textbf{Quality Measure} & \textbf{Best} & \textbf{Default} \\\hline\hline
\multirow{3}{*}{\vspace{-0.3cm}$\begin{matrix}\textbf{Training 400K/4K }\\\textbf{Real dataset}\end{matrix}$}  & \textit{{Threshold}} & 0.4 & 0.5 \\ 
\cline{2-4} & \textit{{$F_1$-score}} & 0.962 & 0.928 \\ \cline{2-4}
 & $\begin{matrix}\textit{{Confusion }} \\ \textit{{matrix}}\end{matrix}$ & \begin{tabular}{c|c} 399817 & 183 \\\hline 459 & 3929 \end{tabular} & \begin{tabular}{c|c} 399877 & 123 \\\hline 1008 & 3380 \end{tabular}\\ \hline\hline
 
 \multirow{3}{*}{\vspace{-0.3cm}$\begin{matrix}\textbf{Training 4K/4K }\\\textbf{Real dataset}\end{matrix}$}  & \textit{{Threshold}} & 0.8 & 0.5 \\ 
\cline{2-4} & \textit{{$F_1$-score}} & 0.919 & 0.793 \\ \cline{2-4}
 & $\begin{matrix}\textit{{Confusion }} \\ \textit{{matrix}}\end{matrix}$ & \begin{tabular}{c|c} 398602 & 1398 \\\hline 197 & 4191 \end{tabular} & \begin{tabular}{c|c} 394172 & 5828 \\\hline 62 & 4326 \end{tabular}\\ \hline
\end{tabular}
}
}
\hfill
\parbox{.48\linewidth}{
\caption{Baseline results using a naive mean-based generator for training. }
\label{table:mean}
\centering
\resizebox{.99\linewidth}{!}{%
\begin{tabular}{|c|c|c|c|}
\hline \textbf{Dataset} & \textbf{Quality Measure} & \textbf{Best} & \textbf{Default} \\\hline\hline
\multirow{3}{*}{\vspace{-0.3cm}$\begin{matrix}\textbf{Training 400K/4K }\\\textbf{Mean-based generation}\end{matrix}$}  & \textit{{Threshold}} & 0.8 & 0.5 \\ 
\cline{2-4} & \textit{{$F_1$-score}} & 0.732 & 0.664 \\ \cline{2-4}
 & $\begin{matrix}\textit{{Confusion }} \\ \textit{{matrix}}\end{matrix}$ & \begin{tabular}{c|c} 396318 & 3682 \\\hline 1894 & 2493 \end{tabular} & \begin{tabular}{c|c} 390060 & 9940 \\\hline 1416 & 2971 \end{tabular}\\ \hline\hline
 
 \multirow{3}{*}{\vspace{-0.3cm}$\begin{matrix}\textbf{Training 4K/4K }\\\textbf{Mean-based generation}\end{matrix}$}  & \textit{{Threshold}} & 0.8 & 0.5 \\ 
\cline{2-4} & \textit{{$F_1$-score}} & 0.601 & 0.583 \\ \cline{2-4}
 & $\begin{matrix}\textit{{Confusion }} \\ \textit{{matrix}}\end{matrix}$ & \begin{tabular}{c|c} 377352 & 22648 \\\hline 839 & 3548 \end{tabular} & \begin{tabular}{c|c} 370528 & 29472 \\\hline 537 & 3850 \end{tabular}\\ \hline
\end{tabular}
}
}
\medskip
\vspace{3ex}

\parbox{.48\linewidth}{
\caption{Performance of synthetic traffic generated by standard WGANs. Results on testing using the best models on training.}
\label{tab:vainilla}
\centering
\resizebox{.99\linewidth}{!}{%
\begin{tabular}{|c|c|c|c|}
\hline \textbf{Dataset} & \textbf{Quality Measure} & \textbf{Best} & \textbf{Default} \\\hline\hline
\multirow{3}{*}{\vspace{-0.3cm}$\begin{matrix}\textbf{Training 400K/4K }\\\textbf{Policy \ref{enum:poli-vainilla-1}) dataset}\end{matrix}$}  & \textit{{Threshold}} & 0.4 & 0.5  \\ 
\cline{2-4} & \textit{{$F_1$-score}} & 0.936 & 0.933 \\ \cline{2-4}
 & $\begin{matrix}\textit{{Confusion }} \\ \textit{{matrix}}\end{matrix}$ & \begin{tabular}{c|c} 399926 & 74 \\\hline 927 & 3461 \end{tabular} & \begin{tabular}{c|c} 399962 & 38 \\\hline 998 & 3390 \end{tabular}\\ \hline\hline
 
 \multirow{3}{*}{\vspace{-0.3cm}$\begin{matrix}\textbf{Training 400K/4K }\\\textbf{Policy \ref{enum:poli-vainilla-2}) dataset}\end{matrix}$}  & \textit{{Threshold}} & 0.8 & 0.5 \\ 
\cline{2-4} & \textit{{$F_1$-score}} & 0.927  & 0.915 \\ \cline{2-4}
 & $\begin{matrix}\textit{{Confusion }} \\ \textit{{matrix}}\end{matrix}$ & \begin{tabular}{c|c} 399449 & 551 \\\hline 701 & 3687\end{tabular} & \begin{tabular}{c|c} 399601 & 399 \\\hline 983 & 3405 \end{tabular}\\ \hline\hline
 
 \multirow{3}{*}{\vspace{-0.3cm}$\begin{matrix}\textbf{Training 4K/4K }\\\textbf{Policy \ref{enum:poli-vainilla-4}) dataset}\end{matrix}$}  & \textit{{Threshold}} & 0.8 & 0.5 \\ 
\cline{2-4} & \textit{{$F_1$-score}} & 0.878 & 0.835 \\ \cline{2-4}
 & $\begin{matrix}\textit{{Confusion }} \\ \textit{{matrix}}\end{matrix}$ & \begin{tabular}{c|c} 399030 & 970 \\\hline 1108 & 3280\end{tabular} & \begin{tabular}{c|c} 396381 & 3619 \\\hline 315 & 4073 \end{tabular}\\ \hline
\end{tabular}
}
}
\hfill
\parbox{.48\linewidth}{
\caption{Performance of synthetic traffic when generators use  custom activation functions in the output. Results on testing using the best models on training.}
\label{table:FA-Leaky}
\centering
\resizebox{.99\linewidth}{!}{%
\begin{tabular}{|c|c|c|c|}
\hline \textbf{Dataset} & \textbf{Quality Measure} & \textbf{Best} & \textbf{Default} \\\hline\hline
\multirow{3}{*}{\vspace{-0.3cm}$\begin{matrix}\textbf{Training 400K/4K }\\\textbf{Policy \ref{enum:poli-vainilla-1}) dataset}\end{matrix}$}  & \textit{{Threshold}} & 0.8 & 0.5  \\ 
\cline{2-4} & \textit{{$F_1$-score}} & 0.649 & 0.555 \\ \cline{2-4}
 & $\begin{matrix}\textit{{Confusion }} \\ \textit{{matrix}}\end{matrix}$ & \begin{tabular}{c|c} 382755 & 17245 \\\hline  241 & 4146 \end{tabular} & \begin{tabular}{c|c} 357385 & 42615 \\\hline 96 & 4291\end{tabular}\\ \hline\hline

 \multirow{3}{*}{\vspace{-0.3cm}$\begin{matrix}\textbf{Training 400K/4K }\\\textbf{Policy \ref{enum:poli-vainilla-2}) dataset}\end{matrix}$}  & \textit{{Threshold}} & 0.8 & 0.5 \\ 
\cline{2-4} & \textit{{$F_1$-score}} & 0.622 & 0.559 \\ \cline{2-4}
 & $\begin{matrix}\textit{{Confusion }} \\ \textit{{matrix}}\end{matrix}$ & \begin{tabular}{c|c} 377693 & 22307 \\\hline 163 & 4224\end{tabular} & \begin{tabular}{c|c} 358897 & 41103 \\\hline 82 & 4305 \end{tabular}\\ \hline\hline

 \multirow{3}{*}{\vspace{-0.3cm}$\begin{matrix}\textbf{Training 4K/4K }\\\textbf{Policy \ref{enum:poli-vainilla-4}) dataset}\end{matrix}$}  & \textit{{Threshold}} & 0.8 & 0.5 \\ 
\cline{2-4} & \textit{{$F_1$-score}} & 0.532 & 0.467 \\ \cline{2-4}
 & $\begin{matrix}\textit{{Confusion }} \\ \textit{{matrix}}\end{matrix}$ & \begin{tabular}{c|c} 345452 & 54548] \\\hline 21 & 4366\end{tabular} & \begin{tabular}{c|c} 298568 & 101432 \\\hline 0 & 4387\end{tabular}\\ \hline
\end{tabular}
}
}

\medskip
\vspace{3ex}
\parbox{.48\linewidth}{
\caption{Performance of synthetic traffic generated by standard WGANs after filtering fake samples by discriminator. Results on testing using the best models on training.}
\label{tab:filtering}
\centering
\resizebox{.99\linewidth}{!}{%
\begin{tabular}{|c|c|c|c|}
\hline \textbf{Dataset} & \textbf{Quality Measure} & \textbf{Best} & \textbf{Default} \\\hline\hline
\multirow{3}{*}{\vspace{-0.3cm}$\begin{matrix}\textbf{Training 400K/4K }\\\textbf{Filtering out fakes}\end{matrix}$}  & \textit{{Threshold}} & 0.6 & 0.5  \\ 
\cline{2-4} & \textit{{$F_1$-score}} & 0.925 & 0.914 \\ \cline{2-4}
 & $\begin{matrix}\textit{{Confusion }} \\ \textit{{matrix}}\end{matrix}$ & \begin{tabular}{c|c} 399511 & 489 \\\hline 767 & 3621 \end{tabular} & \begin{tabular}{c|c} 399221 & 779 \\\hline 722 & 3666 \end{tabular}\\ \hline\hline
\end{tabular}
}
\vspace{3ex}
\caption{Performance of synthetic traffic generated by standard WGANs by sampling generators with elitism among the top $10$ models in training sorted by $F_1$-score. Results on testing.}
\label{tab:FA-linear-top10}
\centering
\resizebox{.99\linewidth}{!}{%
\begin{tabular}{|c|c|c|c|}
\hline \textbf{Dataset} & \textbf{Quality Measure} & \textbf{Best} & \textbf{Default} \\\hline\hline
\multirow{3}{*}{\vspace{-0.3cm}$\begin{matrix}\textbf{Training 400K/4K }\\\textbf{Top $10$ in $F_1$-score}\end{matrix}$}  & \textit{{Threshold}} & 0.4 & 0.5  \\ 
\cline{2-4} & \textit{{$F_1$-score}} &  0.951 & 0.950 \\ \cline{2-4}
 & $\begin{matrix}\textit{{Confusion }} \\ \textit{{matrix}}\end{matrix}$ & \begin{tabular}{c|c} 399800 & 200 \\\hline 608 & 3780 \end{tabular} & \begin{tabular}{c|c} 399832 & 168 \\\hline 658& 3730 \end{tabular}\\ \hline
\end{tabular}
}
}
\hfill
\parbox{.48\linewidth}{
\caption{
Performance of synthetic traffic generated by standard WGANs by sampling generators with elitism among the top $10$ models in training sorted by $L^1$-distance and Jaccard index. Results on testing.}
\label{tab:FA-linear-top10-dKJ}
\centering
\resizebox{.99\linewidth}{!}{%
\begin{tabular}{|c|c|c|c|}
\hline \textbf{Dataset} & \textbf{Quality Measure} & \textbf{Best} & \textbf{Default} \\\hline\hline
\multirow{3}{*}{\vspace{-0.3cm}$\begin{matrix}\textbf{Training 400K/4K }\\\textbf{Top 10 $L^1$-distance}\end{matrix}$}  & \textit{{Threshold}} & 0.8 & 0.5  \\ 
\cline{2-4} & \textit{{$F_1$-score}} & 0.869 & 0.858 \\ \cline{2-4}
 & $\begin{matrix}\textit{{Confusion }} \\ \textit{{matrix}}\end{matrix}$ & \begin{tabular}{c|c} 399491 & 509 \\\hline 1506 & 2882 \end{tabular} & \begin{tabular}{c|c} 398536 & 1464 \\\hline 1094 & 3294 \end{tabular}\\ \hline\hline
 
 \multirow{3}{*}{\vspace{-0.3cm}$\begin{matrix}\textbf{Training 400K/4K }\\\textbf{Top 10 Jaccard index}\end{matrix}$}  & \textit{{Threshold}} & 0.8 & 0.5 \\ 
\cline{2-4} & \textit{{$F_1$-score}} & 0.884  & 0.826 \\ \cline{2-4}
 & $\begin{matrix}\textit{{Confusion }} \\ \textit{{matrix}}\end{matrix}$ & \begin{tabular}{c|c} 399033 & 967 \\\hline 1031 & 3357\end{tabular} & \begin{tabular}{c|c} 396881 & 3119 \\\hline 720 & 3668 \end{tabular}\\ \hline
\end{tabular}
}
}

\end{table*}

Recall that the number of classes in the real dataset are greatly unbalanced, with around $400K$ samples of label $0$ (normal traffic) against $4K$ instances of label $1$ (cryptomining traffic).
This is in perfect agreement with the fact that the previously trained model might tend to predict more frequently class $0$ than class $1$, so there are few false positives. 
For this reason, additionally to the standard training with the whole dataset, we tested the performance after applying a random subsampling strategy that extracts a balanced training dataset with approximately $4K$ instances per class . However, with this strategy, the results of the Random Forest model worsen. The best threshold for the balanced dataset achieves a $F_1$-score significantly smaller than the default value for the unbalanced one (and with the default value, the results are much worse). The confusion matrix in this case inverts the trend, and most of the wrong classified instances are false negative (with a fairly high rate). It is worth mentioning that, however, the number of false negatives drastically decreases with respect to the unbalance setting.

\subsubsection{Na\"ive mean-based generator} \hspace{1cm}
\label{sec:naive-exp}
In this section, we evaluate the performance of the previously described ML classifier when the training dataset is fully substituted with a synthetic dataset generated through a simple mean-based generator. This will serve as a baseline for the upcoming experiments.

This mean-based generator was created by computing the mean and variance of each of the four features of the dataset per class. With these data, a completely new dataset was generated by drawing samples from a multivariate normal centered at the means with diagonal covariance matrix (i.e., each feature is drawn independently) for each class.
Afterwards, we fed the Random Forest model with $300$ classification trees with this synthetic dataset and we analyzed its performance on the (real) test data set (DS2). The results are shown in Table \ref{table:mean}.

Table \ref{table:mean} shows that the performance is much worse with this na\"ively generated dataset. For the best setup (threshold $0.8$ with the unbalanced dataset) the $F_1$-score obtained is $0.732$, significantly smaller than any result got with the real dataset. The situation for the balanced dataset is even worse, with a best $F_1$-score of $0.601$. In both cases, the confusion matrix also shows a concerning phenomenon: the number of false positives is very large, even greater than the number of true positives. 

Since the generating method for the dataset is intrinsically stochastic (new instances are generated by sampling a random vector), in Figure \ref{fig:mean} we plot the histogram of the obtained $F_1$-scores for several runs of the generative method. None of the tested instances were able to reach a $F_1$-score greater than $0.76$. This strengthen the values shown in Table \ref{table:mean}, showing that they are actually statistically consistent and different runs of the generative method return similar results.

These results evidence that the na\"ive mean-based generator is not a good approach for generating a synthetic dataset when the data distributions are not easily separable as it happens in our scenario (Section \ref{sec:problem-setting}). The means of the selected features are so close that the generated features of different classes overlap and do not capture the real distribution of the data. Hence, these results point out that a much subtler method of generation is required to obtain a compelling performance.

\subsubsection{Standard GAN}\label{sec:results-standard-GAN}

In this section, we discuss the performance results attained by a simple WGAN with linear activation functions in the output layer. No extension of the method is implemented. This kind of models are sometimes referred to as `vainilla GANs' in the literature due to their simplicity.

Recall from subsection \ref{subsec:architecture} that two WGANs were trained to generate samples corresponding to label $0$ (normal traffic) and label $1$ (cryptomining traffic). 
Using the variables and intervals defined in table \ref{table:hyperparameters}, the hyperparameter setup for the two WGANs was conducted through a random search by screening the $F_1$-score obtained by a nested Random Forest model that evaluated the marginal quality of the generator in testing with respect to the corresponding label (subsection \ref{subsec:experimental-setup}).

The evaluation of the marginal quality of each hyperparameter configuration was done at the end of each mini-batch training phase, computing the marginal $F_1$-score and saving the stage of the WGAN for later use. 
Hence, instead of a single trained WGAN, we got many different WGAN models with  the same hyperparameter setup but at different stages of the training. Thus, each WGAN hyperparameter configuration generated as many $F_1$-scores as mini-batch training steps.
Finally, for each type of traffic, we selected the WGAN configuration that obtained the best $F_1$-score in any of its mini-batches.

For normal traffic, the set of hyperparameters that produced the best performing WGAN was as follows: 
\begin{itemize}
    \item Generator. Architecture: [123,200,500,3000,500,4],  output activation: linear,  latent vector size: 123, multipoint single-class embedding: False, noise for latent vector: Normal (0,5), batch normalization: True, Percentage of tanh: 15\%
    \item Discriminator. Architecture: [4,380,800,600,177,23,1],
output filtering: False, noise in input (real and fakes): N(0,0.02), noise in fakes: N(0,0), ratio label change: 0, batch normalization: True, regularization 0.02, dropout:0.1, learning rate RMS-Prop:0.001.
\end{itemize}

For cryptomining connections, the set of hyperparameters that produced the best performing WGAN was as follows: 
\begin{itemize}

    \item Generator. Architecture: [123,600,3000,1000,4],  output activation: linear,  latent vector size: 123, multipoint single-class embedding: False, noise for latent vector: uniform (0,3.5), batch normalization: True, Percentage of tanh: 5\%
    \item Discriminator. Architecture: [4,280,903,500,23,1],
output filtering: False, noise in input (real and fakes): N(0,0.01), noise in fakes: N(0,0), ratio label change: 0, batch normalization: True, regularization 0.05, dropout:0.15, learning rate RMS-Prop:0.001.
\end{itemize}

All WGANs were trained at least 1000 mini-batch steps for label "0" and 1400 for label "1". Recall that we have around 400,000 samples of label "0" and only 4,000 of label "1" in DS1 and DS2 (training and testing datasets).
The size of label "1" mini-batches was configured 10 times less than label "0" mini-batches, and therefore, one epoch of label "0" implied 500 mini-batch train steps, and one epoch of label "1" consisted of only 50 mini-batch train steps.  This is the reason label "1" WGANs were trained with more mini-batches than label "0" WGANs in the same period of time.   

Having selected the best performing WGAN configurations for each type of traffic, and in order to compare the quality of the synthetic traffic with respect to the real traffic, one generator of each type of traffic was chosen from all mini-batch models previously saved during training. Using this pair of generators, we obtained a combination of samples of the two types of traffic that formed a fully synthetic dataset. The synthetic dataset  was used to feed the training of a nested Random Forest classifier (with $300$ trees) that was subsequently tested with real data (DS2 dataset). 
The selection of a pair of generators and the number of samples produced by each of them was done by using the following three policies:
\begin{enumerate}
    \item\label{enum:poli-vainilla-1} An unbalanced dataset (with $400K/4K$ instances) is generated by picking at random {one} model for each label among the partially trained models.
    \item\label{enum:poli-vainilla-2} An unbalanced dataset (with $400K/4K$ instances) is generated by picking at random {two} models for each label. The instances of the synthetic dataset were obtained by mixing the outputs of the two chosen generative models, in the expectancy of increasing the variety and diversity of the synthetic dataset.
    \item\label{enum:poli-vainilla-4} A {balanced} dataset (with $4K/4K$ instances) is generated by picking at random {one} model for each label.
\end{enumerate}

For each policy, the selection of the pair of generators (one for each type of traffic) was drawn 20 times uniformly at random from the generators of each WGAN configuration. The $F_1$-scores obtained at the end of these 20 experiments are shown in Figure \ref{fig:vainilla} (Appendix I). It is worth noting that sampling pairs of generators allow us to study the statistical distribution of the standard WGAN quality metrics without evaluating all possible combinations of generators.

The results reached by the best model in these experiments are shown in Table \ref{tab:vainilla}. The quality measures point out that the best results were obtained when we applied policy \ref{enum:poli-vainilla-1}), with a slightly better performance compared to the results got when real data was used (Table \ref{table:real}) and much better performance than for the na\"ive mean-based generator (Table \ref{table:mean}). In contrast, policy \ref{enum:poli-vainilla-2})  reached a lower performance than the single-model policy. However, if we look at the histograms depicted in Figure \ref{fig:vainilla} (subplots (a) and (b)), we observe that even though the best model is obtained with policy \ref{enum:poli-vainilla-1}), the datasets obtained by mixing two models, as provided by policy \ref{enum:poli-vainilla-2}), tend to be less dependent on the sampled models, with a more uniform distribution of the $F_1$-scores. 

This points out that the variety and diversity generated by using policy \ref{enum:poli-vainilla-1}) tends to generate a large amount of information in the dataset, which can be exploited by the nested ML model. 
Should we increase even more the variability of the dataset by mixing two models as in policy \ref{enum:poli-vainilla-2}), the results tend to be more consistent between executions, even though the best results are slightly worse due to the added spurious noise. 
Moreover, in the line of the results obtained in subsection \ref{sec:experiments-real}, the policy \ref{enum:poli-vainilla-4}) with a generated balanced dataset did not reach compelling results.

\subsubsection{Evolution of the training process for standard GAN}
\label{subsec:gan_vainilla_results}

To analyze more deeply the results of Section \ref{sec:results-standard-GAN}, in this subsection we shall evaluate the evolution of the different quality measures of the best standard WGAN model throughout its training process.

\begin{figure*}[!t]

\begin{subfigure}[t]{.48\textwidth}
\centering
\includegraphics[width=1\linewidth]{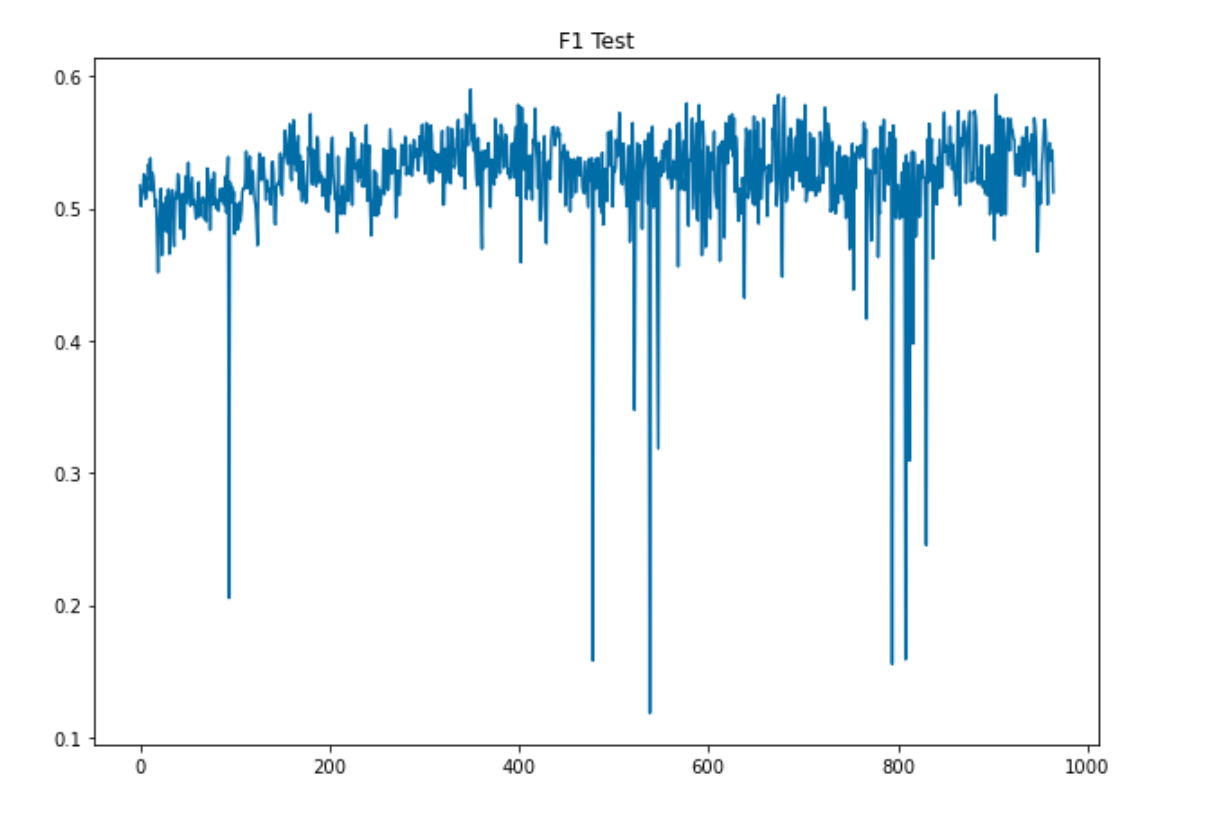} 
\caption{$F_1$-score on testing}
\label{fig:vainilla-f1}
\end{subfigure}
\begin{subfigure}[t]{.48\textwidth}
\centering
\includegraphics[width=1\linewidth]{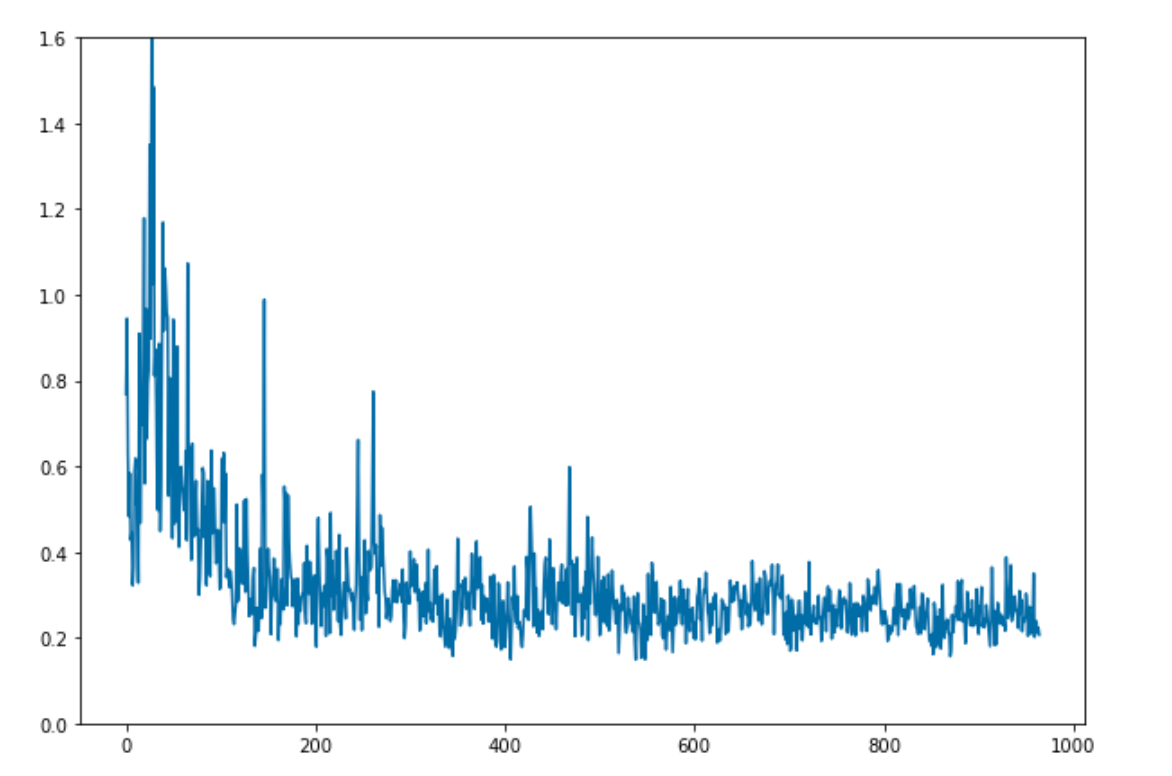} 
\caption{$L^1$ distance}
\label{fig:vainilla-kolmo}
\end{subfigure}

\medskip

\begin{subfigure}[t]{.48\textwidth}
\centering
\includegraphics[width=1\linewidth]{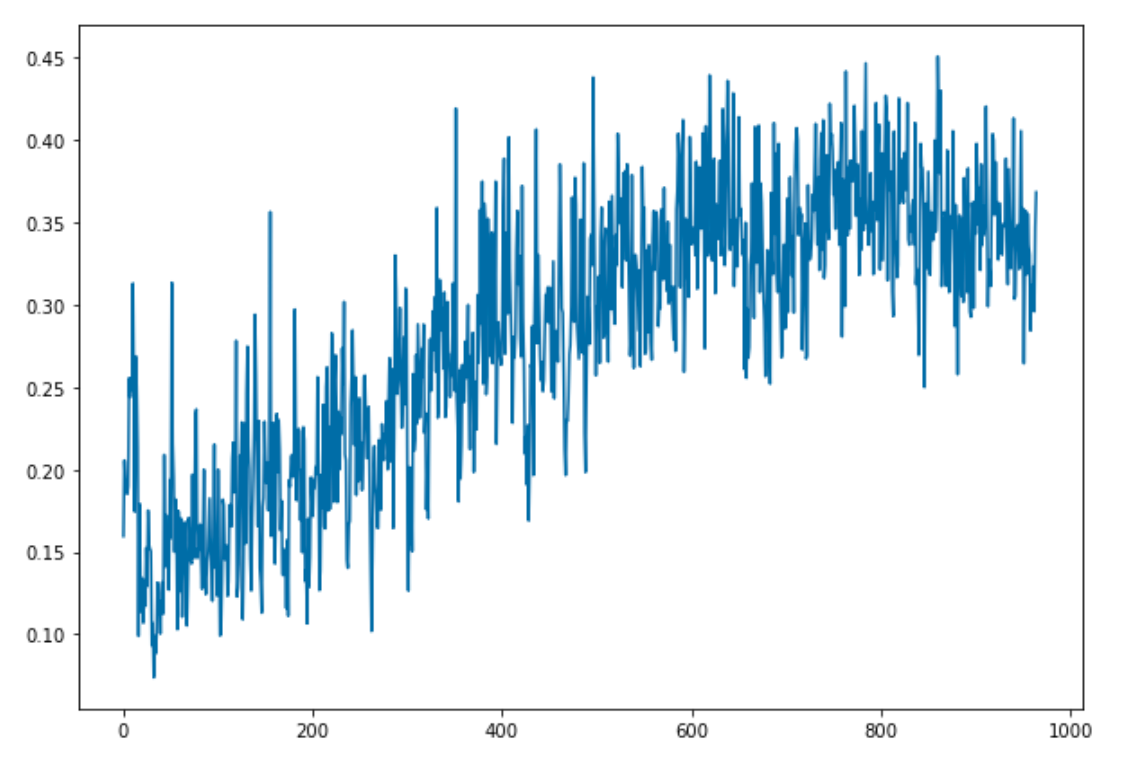} 
\caption{Jaccard index}
\label{fig:vainilla-jac}
\end{subfigure}
\begin{subfigure}[t]{.48\textwidth}
\centering
\includegraphics[width=1\linewidth]{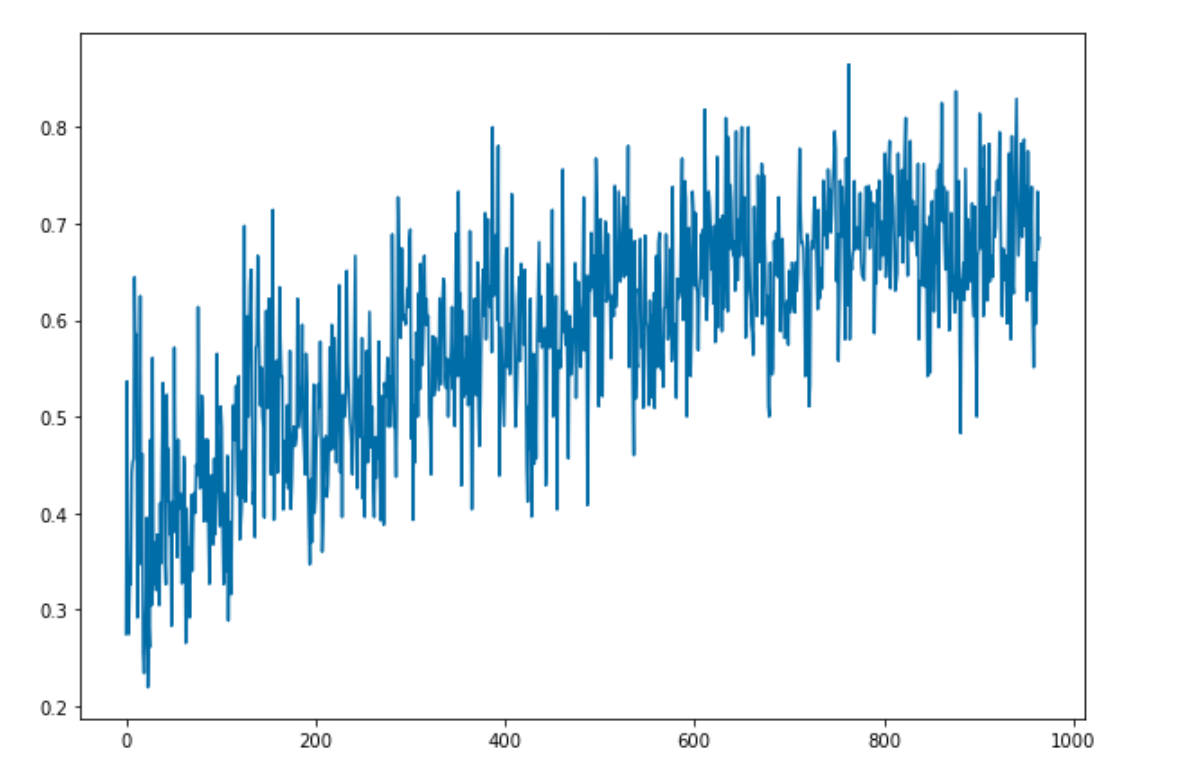} 
\caption{Jaccard index from percentile 1}
\label{fig:vainilla-jac-1}
\end{subfigure}

\caption{Evolution of $F_1$-score on testing, $L^1$ distance and Jaccard index using GAN generator for label 0. The $x$-axis represents the GAN training epochs.}
\label{fig:distances-vainilla}
\end{figure*}

\begin{figure*}[!t]

\begin{subfigure}[t]{.48\textwidth}
\centering
\includegraphics[width=1\linewidth]{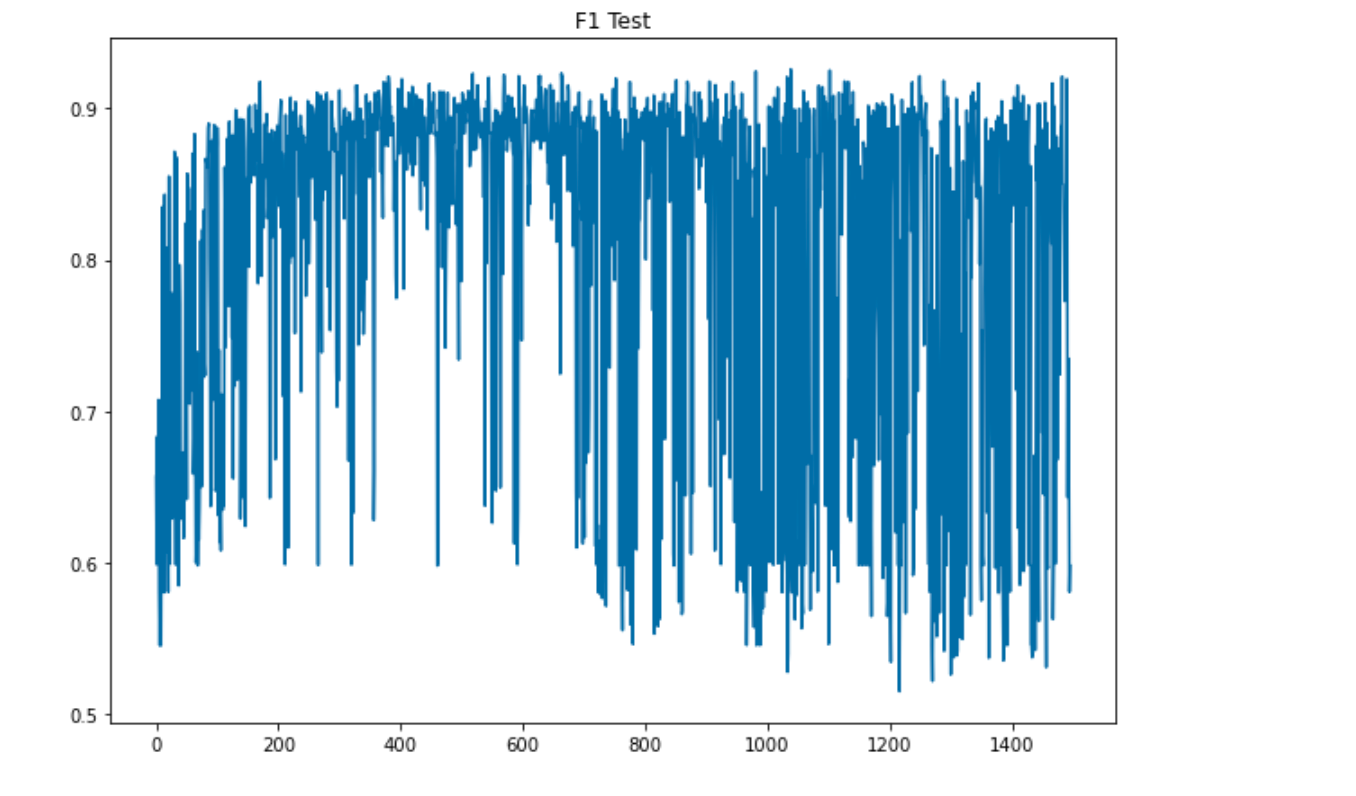} 
\caption{$F_1$-score on testing}\label{fig:vainilla-f1-et1}
\end{subfigure}
\begin{subfigure}[t]{.48\textwidth}
\centering
\includegraphics[width=1\linewidth]{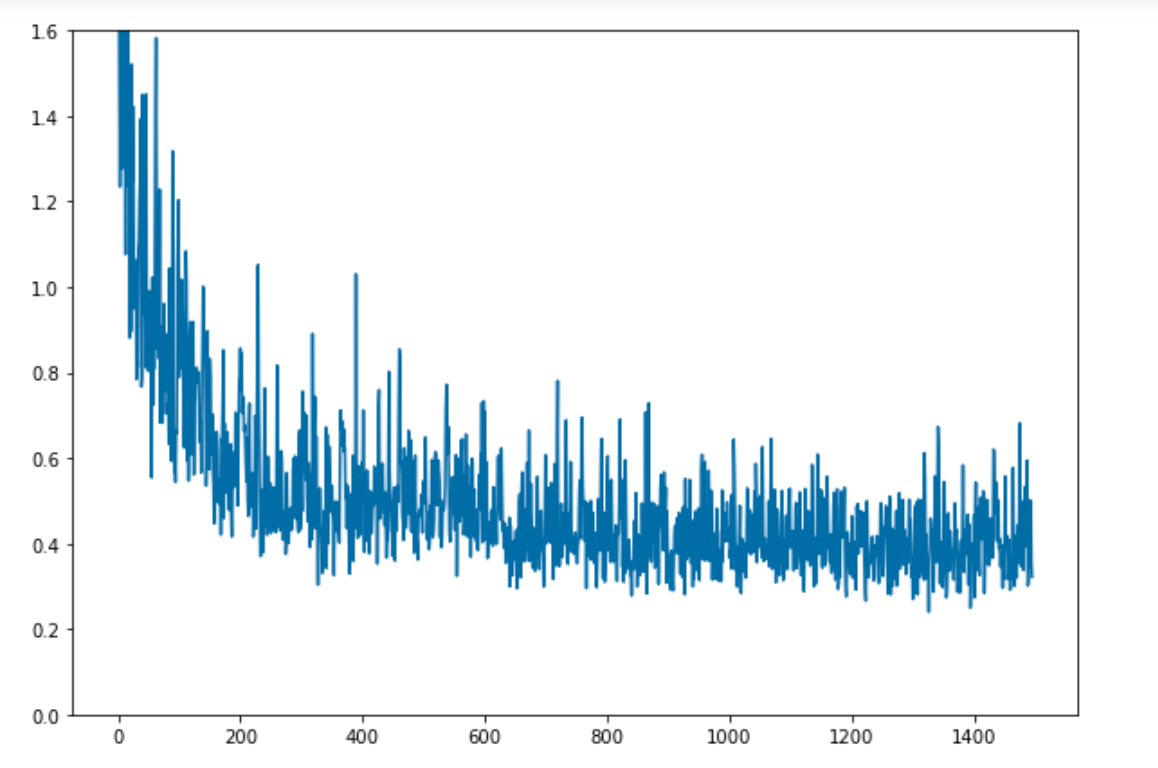} 
\caption{$L^1$ distance}\label{fig:vainilla-kolmo-et1}
\end{subfigure}

\medskip

\begin{subfigure}[t]{.48\textwidth}
\centering
\includegraphics[width=1\linewidth]{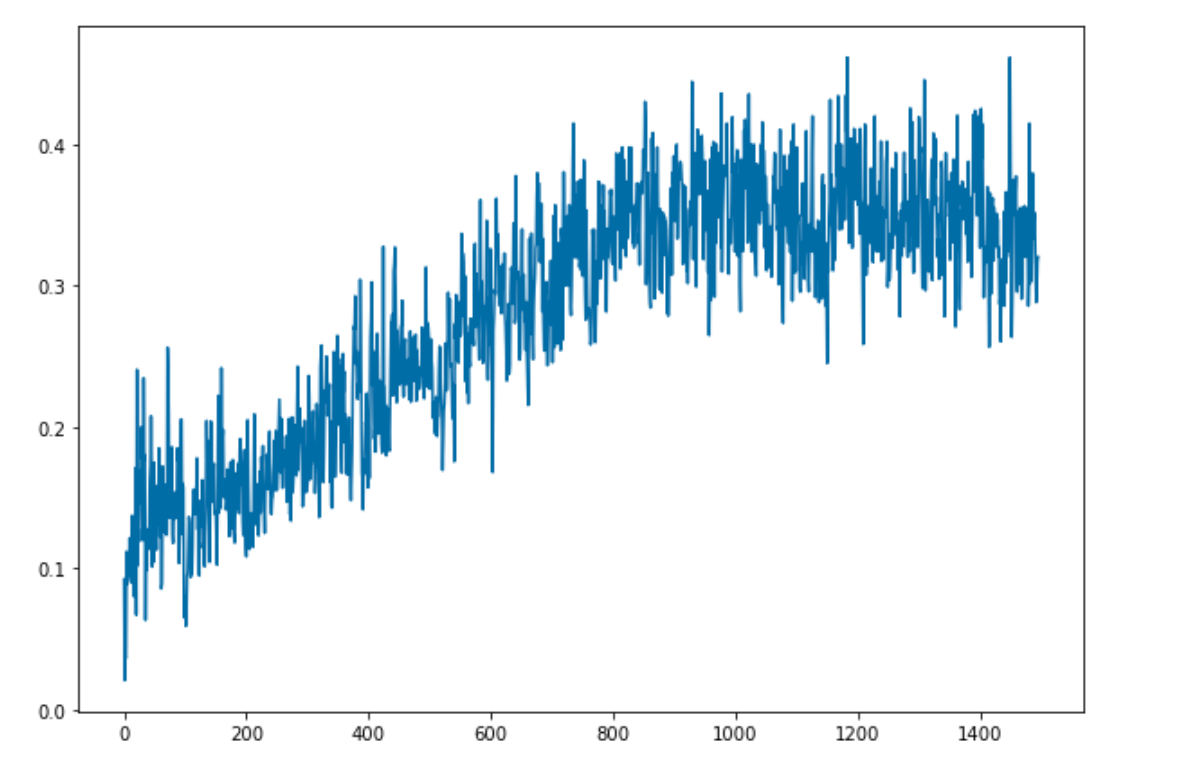} 
\caption{Jaccard index}\label{fig:vainilla-jacc-et1}
\end{subfigure}
\begin{subfigure}[t]{.48\textwidth}
\centering
\includegraphics[width=1\linewidth]{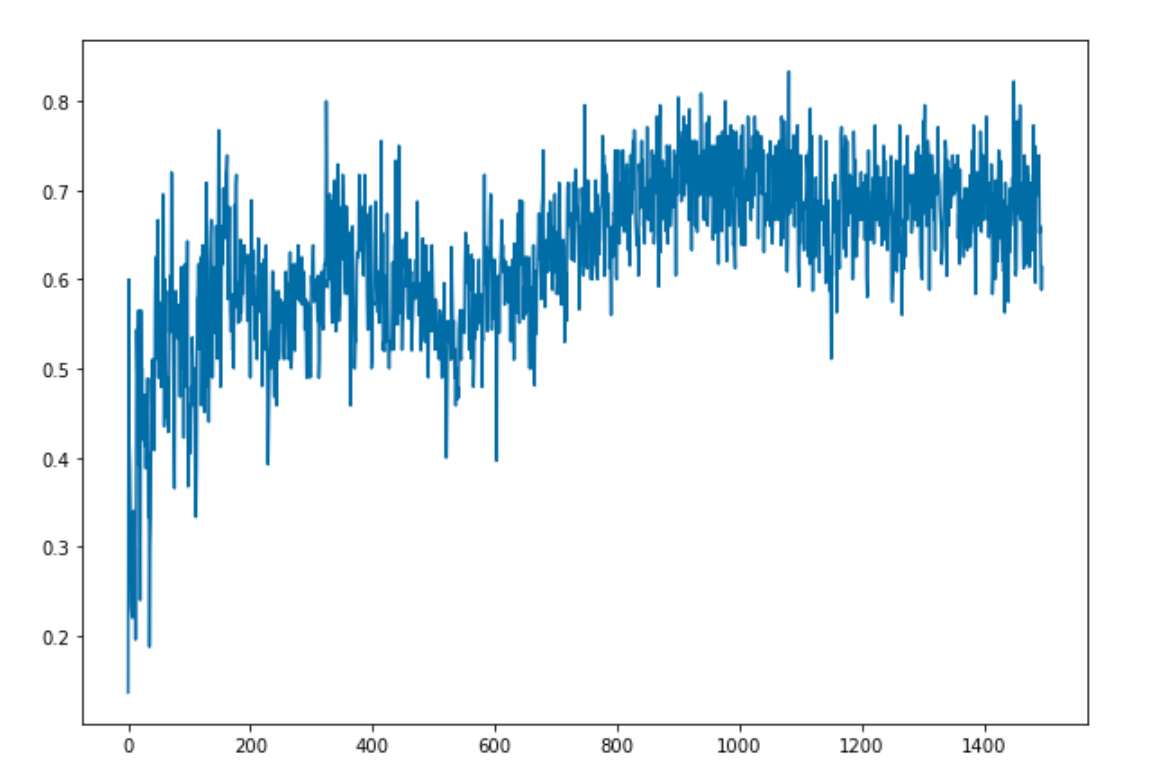} 
\caption{Jaccard index from percentile 1}\label{fig:vainilla-jacc-1-et1}
\end{subfigure}

\caption{Evolution of $F_1$-score on testing, $L^1$ distance and Jaccard index using GAN generator for label 1. The $x$-axis represents GAN training epochs.}
\label{fig:distances-vainilla-et1}
\end{figure*}

In Figures \ref{fig:distances-vainilla} and \ref{fig:distances-vainilla-et1}, we show the evolution of the different metrics described in Section \ref{sec:metrics} for the different training epochs of the best WGAN model for label $0$ and label $1$, respectively. In particular, Figure \ref{fig:vainilla-f1} shows the evolution of the $F_1$-score in testing when generated data for the label $0$ (normal traffic) is mixed with real data for label $1$ (cryptomining traffic), while Figure \ref{fig:vainilla-f1-et1} depicts the evolution of the $F_1$-score when label $1$ is generated and real data is used for label $0$. As we can observe, for both labels the obtained results are quite consistent along the training process (beware of the scale of the plots), with some marked drops for label $0$ that are rapidly recovered, probably due to drastic changes in the generator network.

However, this constant tendency in the $F_1$-score is in sharp contrast with the evolution indicators described in Section \ref{sec:metrics} of statistical nature, such as the $L^1$-distance (Figures \ref{fig:vainilla-kolmo} and \ref{fig:vainilla-kolmo-et1}) and the Jaccard index of the support (Figures \ref{fig:vainilla-jac} and \ref{fig:vainilla-jacc-et1}) with respect to the original distribution. For these statistical coefficients, we observe that a longer training usually leads to generated samples of better quality (smaller $L^1$ distance and larger Jaccard index), as predicted by the theoretical convergence results for GANs. This is a very interesting observation, since it evidences that a better performance of the generated data in a nested ML is not directly related with a better fit with the original distribution. In this manner, classical measures of the goodness of fit are not good estimators of the information contained in the generated data, as can be exploited by a nested ML model. 
This observation leads us to conjecture that these metrics are not reflecting some quality parameter indicating that the synthetic data lack some essential information that the real data does.

Analysing $F_1$-scores of each type of traffic, it can be observed that label "0" WGAN generates synthetic data that performs worse than label "1" WGAN when the synthetic data is used to train a Random Forest classifier. Label "0" synthetic traffic does not achieve a $F_1$-score greater than $0.58$ but label "1" synthetic traffic obtains $F_1$-scores greater than $0.9$. Recall that label "0" traffic is a complex mixture of web, video, shared-folder and other protocols and on the contrary, label "1" traffic is a more homogeneous traffic generated by four types of cryptomining protocols. We conjecture that the complexity of the former is more difficult to be learnt and generated by WGANs than the latter.


\begin{figure*}[!t]
\centering
\centering
\begin{subfigure}[t]{.195\textwidth}
\includegraphics[width=1\linewidth]{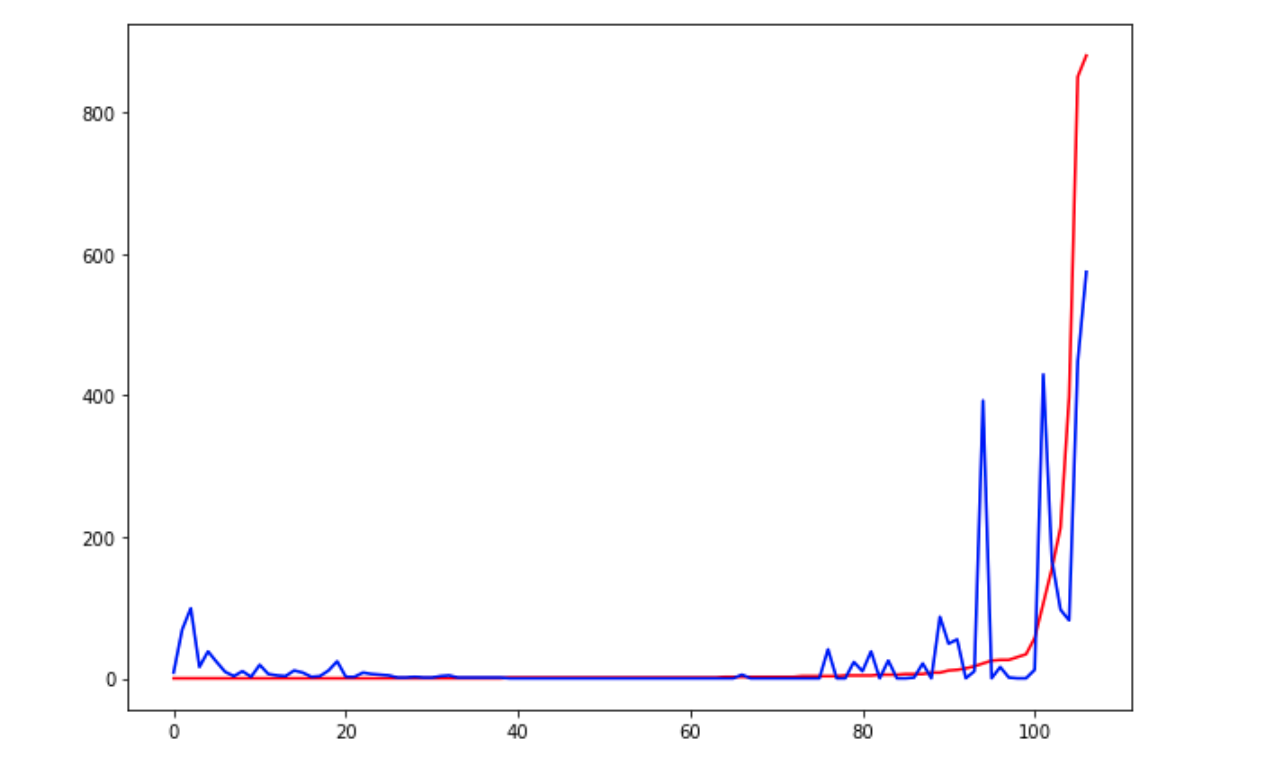} 
\caption{Label 0. Epoch 1}\label{fig:distr-l0-1}
\end{subfigure}
%
\begin{subfigure}[t]{.195\textwidth}
\includegraphics[width=1\linewidth]{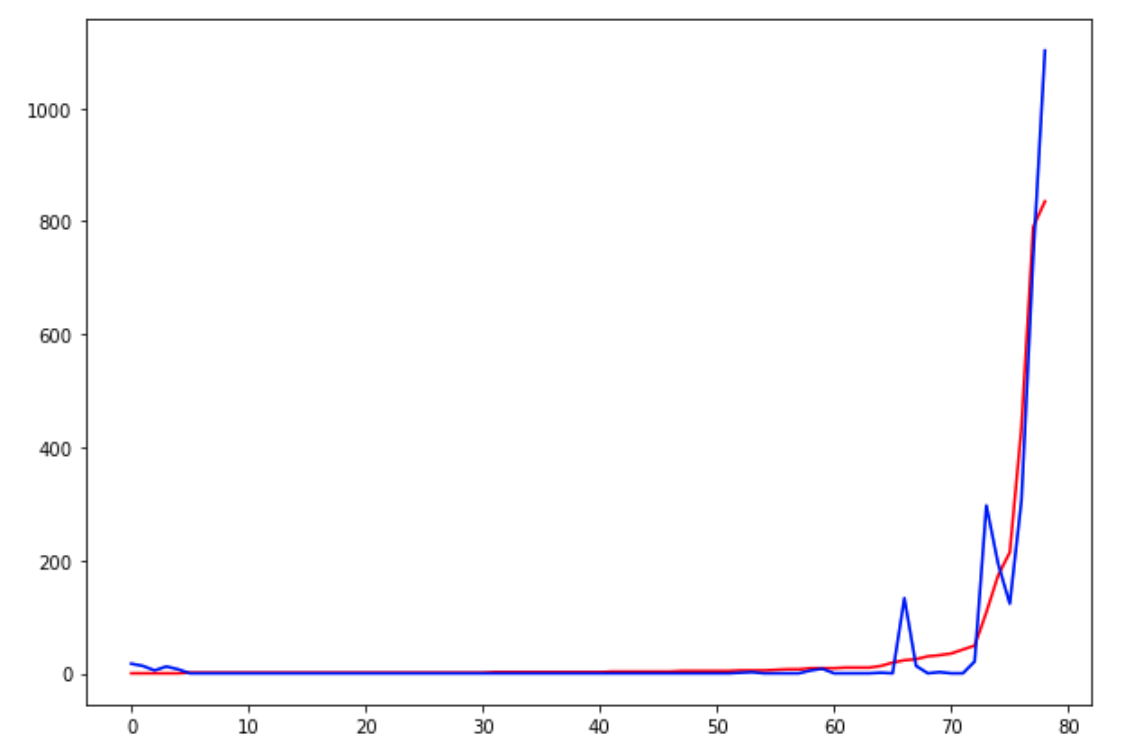} 
\caption{Label 0. Epoch 5}\label{fig:distr-l0-5}
\end{subfigure}
%
%
\begin{subfigure}[t]{.195\textwidth}
\includegraphics[width=1\linewidth]{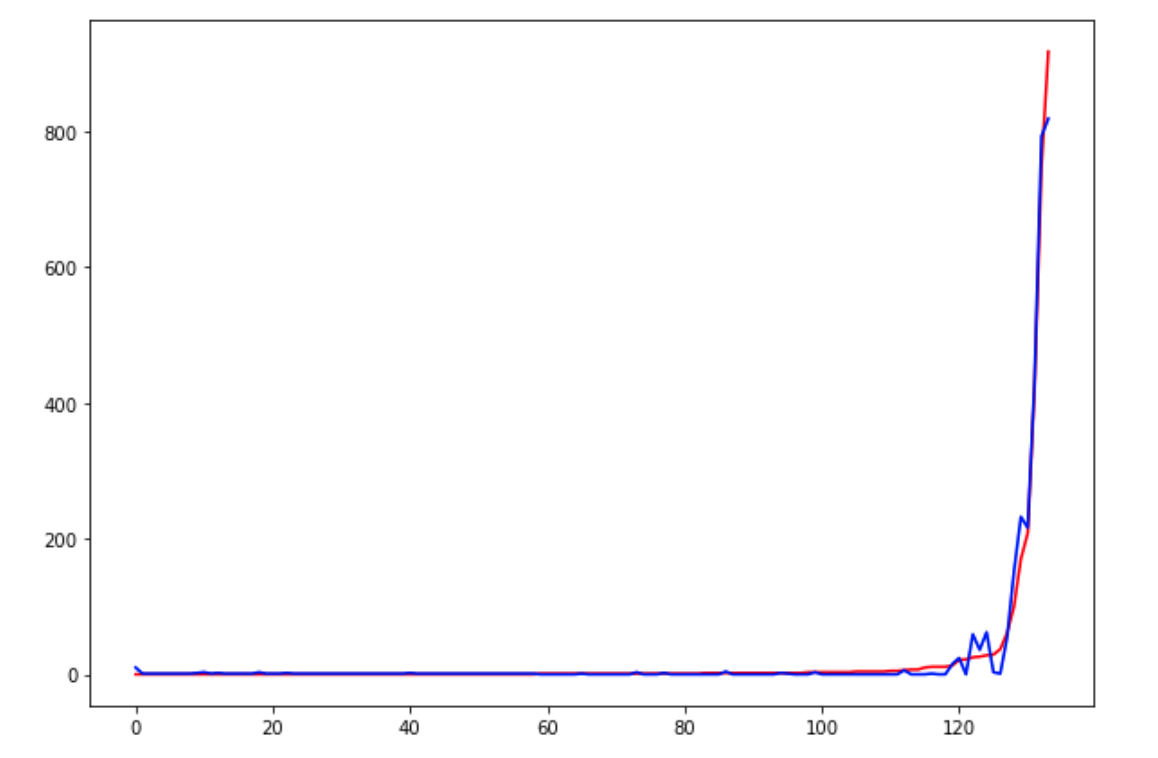} 
\caption{Label 0. Epoch 200}\label{fig:distr-l0-200}
\end{subfigure}
\begin{subfigure}[t]{.195\textwidth}
\includegraphics[width=1\linewidth]{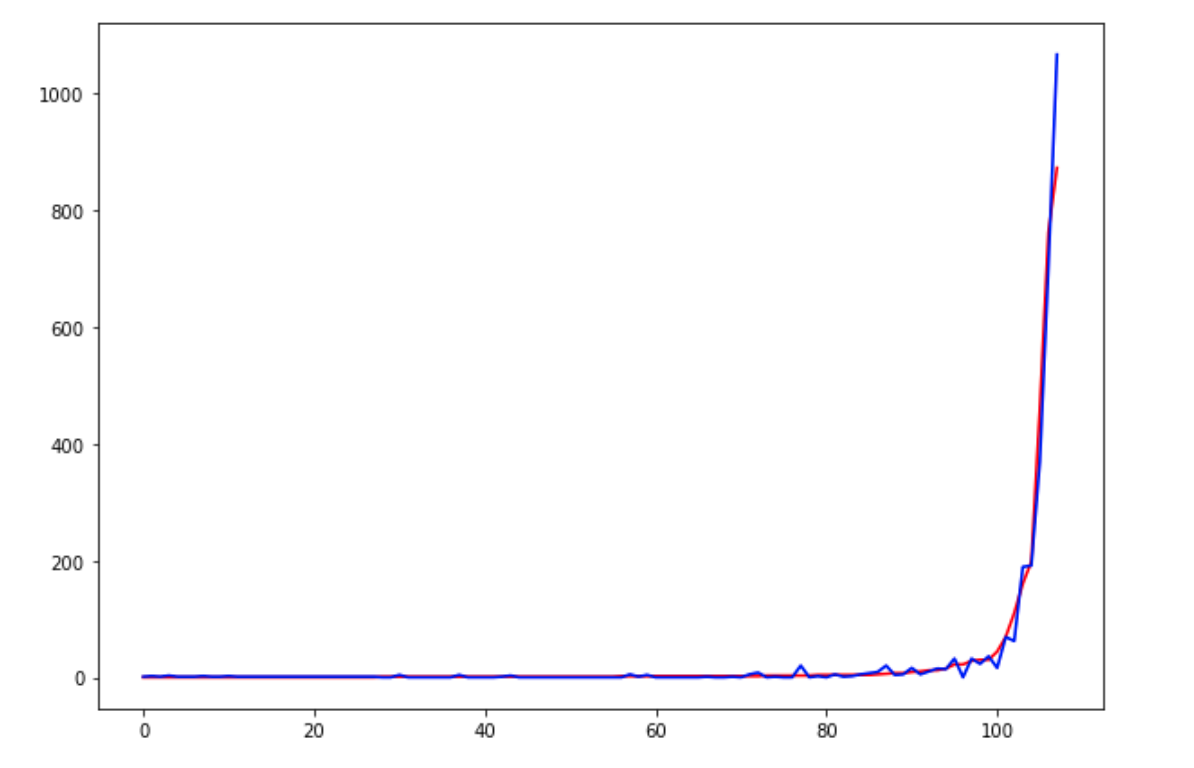} 
\caption{Label 0. Epoch 800}\label{fig:distr-l0-800}
\end{subfigure}
\begin{subfigure}[t]{.195\textwidth}
\includegraphics[width=1\linewidth]{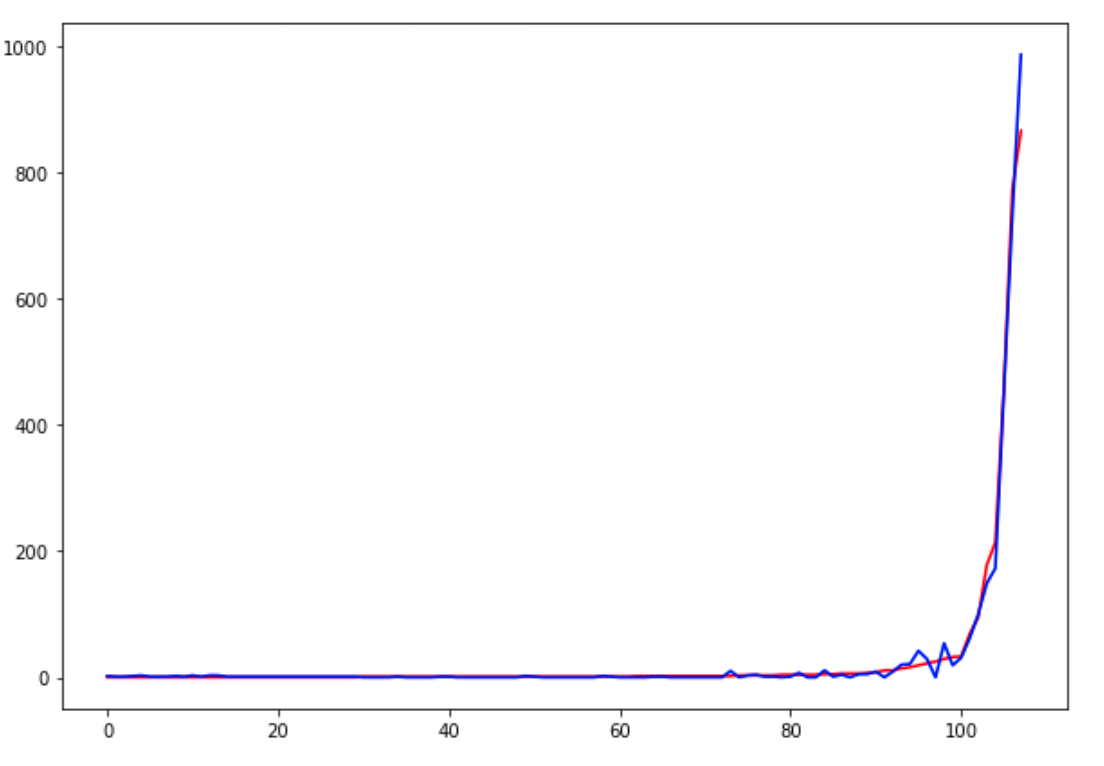} 
\caption{Label 0. Epoch 1000}\label{fig:distr-l0-1500}
\end{subfigure}

\medskip

\begin{subfigure}[t]{.195\textwidth}
\includegraphics[width=1\linewidth]{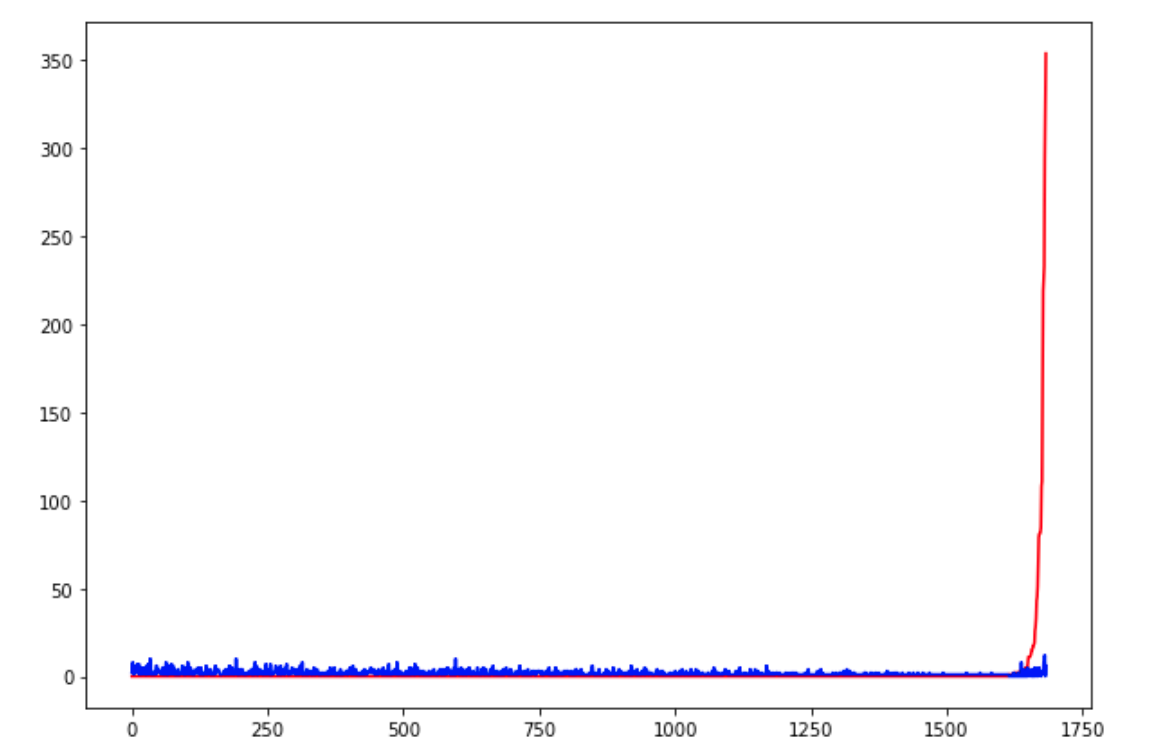} 
\caption{Label 1. Epoch 1}\label{fig:distr-l1-1}
\end{subfigure}
%
\begin{subfigure}[t]{.195\textwidth}
\includegraphics[width=1\linewidth]{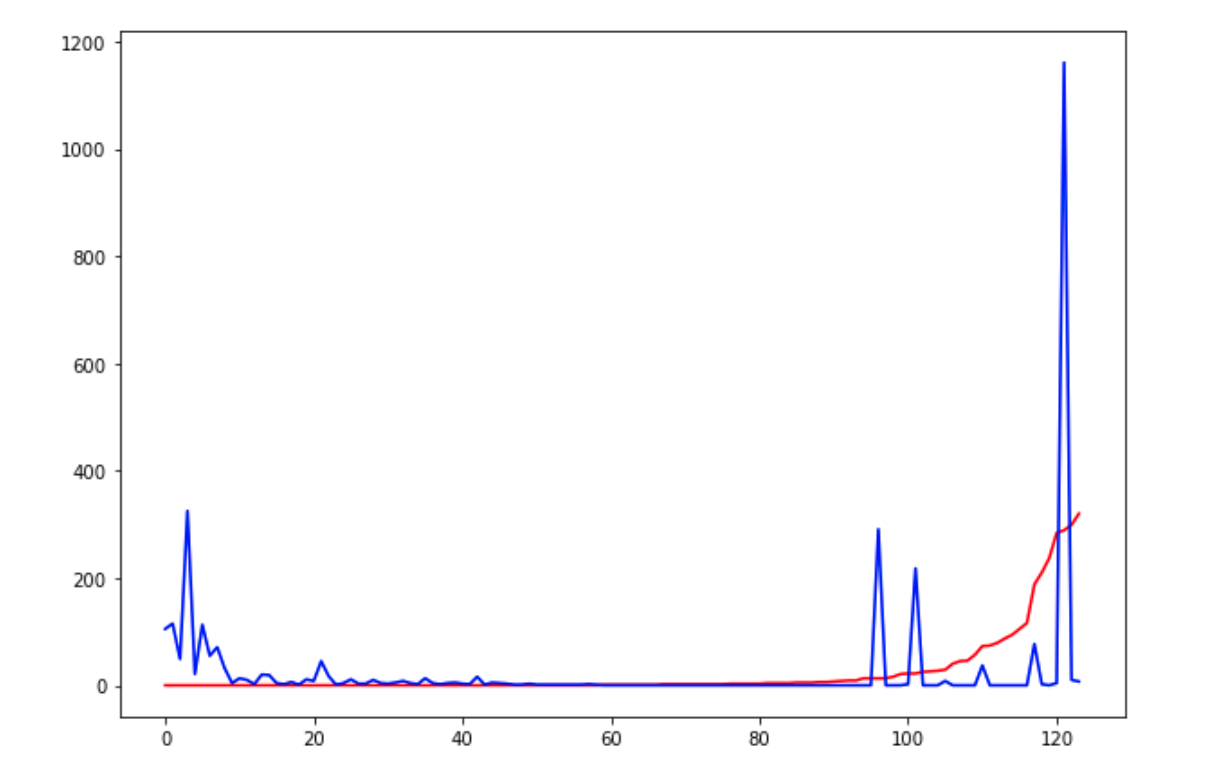} 
\caption{Label 1. Epoch 5}\label{fig:distr-l1-5}
\end{subfigure}
%
%
\begin{subfigure}[t]{.195\textwidth}
\includegraphics[width=1\linewidth]{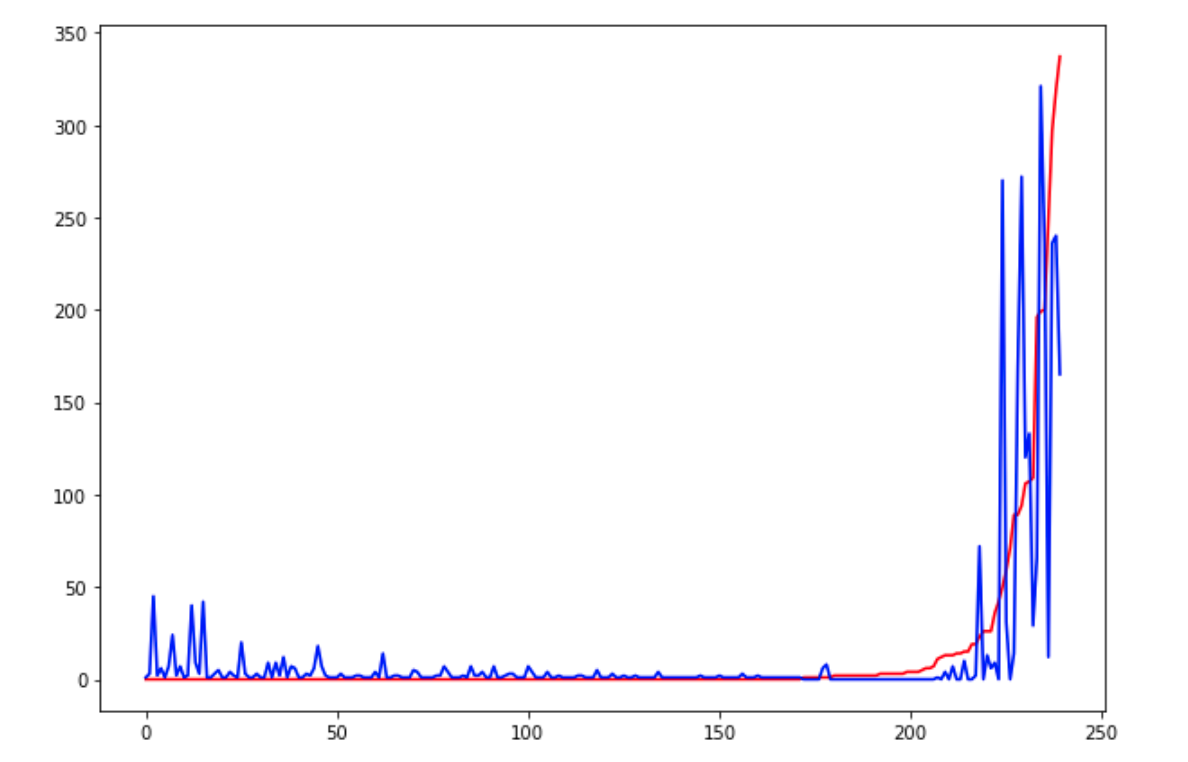} 
\caption{Label 1. Epoch 200}\label{fig:distr-l1-200}
\end{subfigure}
%
\begin{subfigure}[t]{.195\textwidth}
\includegraphics[width=1\linewidth]{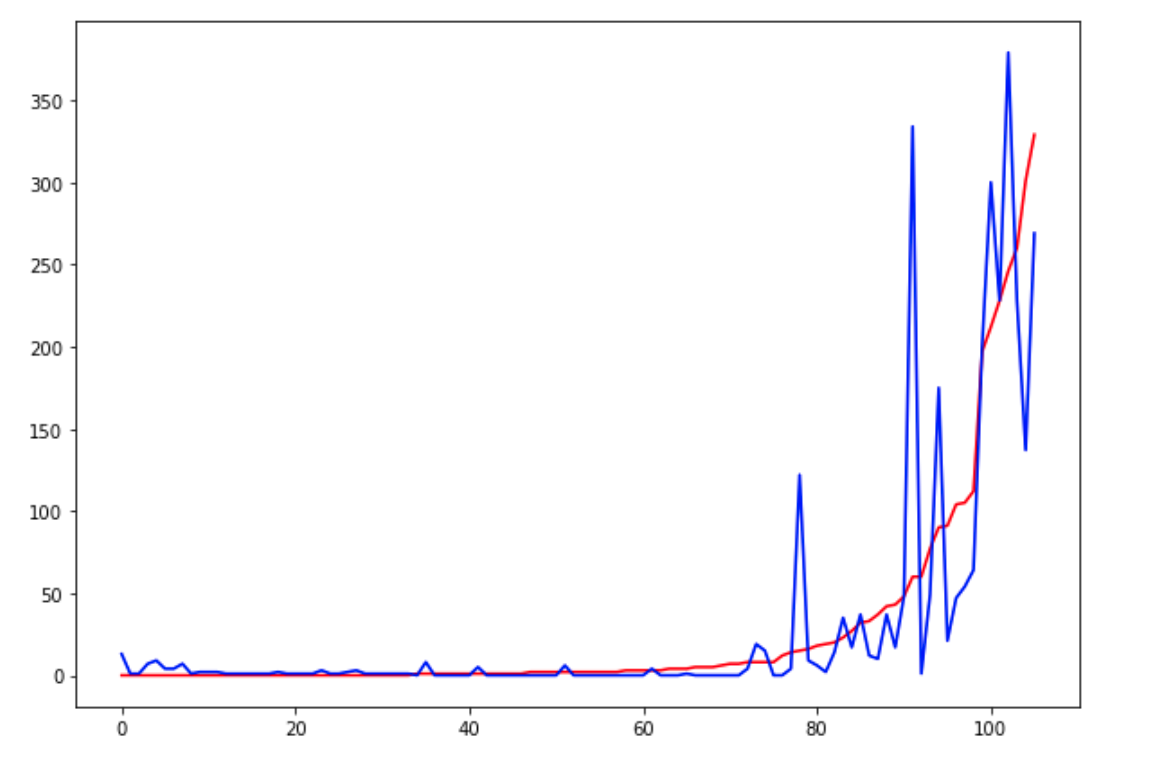} 
\caption{Label 1. Epoch 800}\label{fig:distr-l1-800}
\end{subfigure}
\begin{subfigure}[t]{.195\textwidth}
\includegraphics[width=1\linewidth]{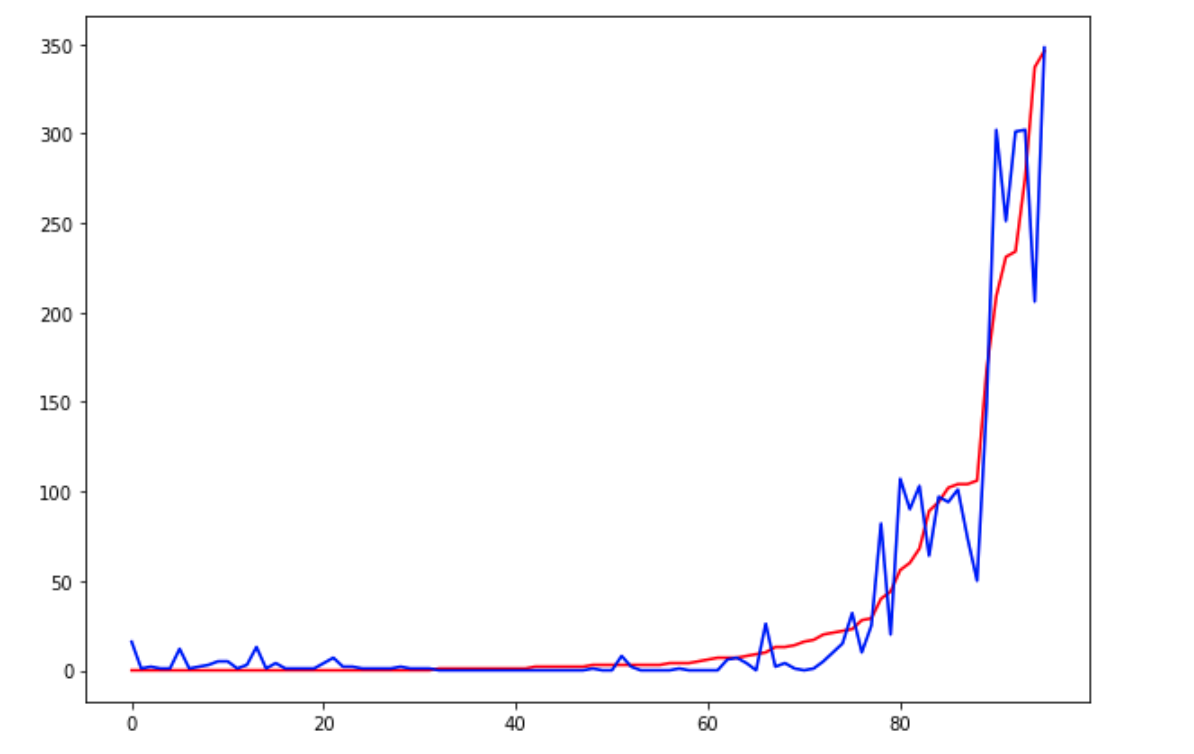} 
\caption{Label 1. Epoch 1500}\label{fig:distr-l1-1500}
\end{subfigure}
\caption{Comparison of synthetic (blue) and real (red) data distributions using GAN generators for label 0 (\ref{fig:distr-l0-1}, \ref{fig:distr-l0-5}, \ref{fig:distr-l0-200}, \ref{fig:distr-l0-800} and \ref{fig:distr-l0-1500}) and label 1 (\ref{fig:distr-l1-1}, \ref{fig:distr-l1-5}, \ref{fig:distr-l1-200}, \ref{fig:distr-l1-800} and \ref{fig:distr-l1-1500}) in different epochs (1, 5, 200, 800 and 1500). The $4$-dimensional vector has been flattened into by sorting by frequency in ascending order on the $x$-axis.}
\label{fig:comparison-vainilla}
\end{figure*}

To strengthen these ideas, in Figure \ref{fig:comparison-vainilla} we compare  the histogram of real (red color) and synthetic (blue color) data  along the training process. 
In X axis we order the intervals $s_0 < s_1 < \ldots < s_\ell$ by $h_X(s)$ values (as defined in equation \ref{eq:1}) of the real data from smallest to largest.
As we can observe, initially the fit to the target distribution is very poor but rapidly the GANs are able to detect and replicate the most frequent values. 
It is worth noting that during the first epochs, WGANs generate a significant number of nonexistent values (left side of the curves) that tend to disappear as training progresses. 
For large times, the real and synthetic distributions are quite similar in agreement with the decrease of the $L^1$-distance and the increase of the Jaccard index.


To finish this section, it is worth mentioning that, even though the theoretical results predict an asymptotic convergence of the generated data to the original distribution, Figures \ref{fig:vainilla-kolmo} and \ref{fig:vainilla-jac} seem to point out that this evolution tends to stuck. 
This stagnation of the quality measures may be caused by the complexity of the original data, which follow very complicated distributions with domain constrains such as non-negativity of discreteness, in contrast with the usual graphical data that is usually taken as input for GANs. However, further research is needed to clarify these issues. 

\section{Effects of improvements and variants}
\label{sec:improvements}

In this section, we shall discuss the effect of the different variants of GANs introduced in Section \ref{sec:proposed_model} for the problem of generating synthetic traffic data. The setting will be the same as in Section \ref{sec:results-standard-GAN}, in which we shall use the performance of a nested ML model as a quantitative metric to compare the variants with the standard WGAN architecture.

\subsection{Custom activation function}
\label{sec:custom-act-func}

In this section, we will evaluate the performance of the GAN after applying the improvement described in Section \ref{sec-sub:custom-ouput}. Roughly speaking, recall that this improvement consists in changing the activation function of the output layer of the generator network to a Leaky-ReLU function with a small slope. Thanks to this plug-in, we are able to preserve the domain semantic constraints such as non-negative values for counters.
Recall that linear activation functions tend to produce bell-shaped data distributions and so, negative values can be generated (left side of the data distribution curve). If the output variable in this linear activation function is a counter or an accumulator, we would be generating nonexistent values. 

The obtained results are shown in Table \ref{table:FA-Leaky} and Figure \ref{fig:FA}. As we can observe, the results are clearly worse than the ones obtained with the standard WGAN in subsection \ref{subsec:gan_vainilla_results}.
In contrast to the observation of subsection \ref{sec:results-standard-GAN}, the performance of the nested ML model seems to be independent of applying sampling policy \ref{enum:poli-vainilla-1}) or \ref{enum:poli-vainilla-2}). When the synthetic dataset is balanced using policy \ref{enum:poli-vainilla-4}), the performance slightly gets worse.

However, it is remarkable from these results that the number of false negatives suffered by the ML model drastically decreases in this case in comparison with the standard GAN. The rationale behind this fact is that, even though the global quality of the data is not as good as with the standard GAN, the preservation of the domain constraints allows the system to be more aggressive in the distinction of cryptomining traffic (obviously, with the drawback of a large rate of false positives). This output may be very useful in those scenarios in which the skipping a flow of cryptomining traffic is very penalized (e.g., due to security issues) but getting a high rate of false positives is not so serious (say, because the only consequence is that the connection is artificially restarted). In these scenarios, the solution based on customized activation functions would be the choice.

In addition, the use of custom activation functions in the generator output layer will be appropriate if our goal is not only to use synthetic data to train a ML-based classifier (e.g., a cryptomining attack detector), but to obtain synthetic data that can be used in other applications in which it is crucial that the data do not contain any nonexistent value (e.g. counters with negative values).  

\subsection{Discriminator as quality assurance}
\label{sec:discr-quality-assurance}

Another interesting approach to obtain high-quality generated features is to use the discriminator network of the GAN as a quality assessor. To be precise, after training the GAN, instead of using all samples synthesized by the generator network, we add to the generated dataset only those that were classified as real samples by the discriminator agent ($D(x) > 0$), while the samples judged as fake ($D(x) \leq 0$) are ruled out.

In this manner, only those samples that were competitive enough to cheat the discriminator were selected. Obviously, this requires to generate significantly more examples than the strictly needed since most of them will be filtered out. However, notice that the generative process corresponds to a feedforward procedure and thus is quite fast, so the required time is not significantly larger.

To test this idea, an experiment was carried out following the sampling policy \ref{enum:poli-vainilla-1}. The results of this analysis are shown in Table \ref{tab:filtering} and Figure \ref{fig:vainillaPos_400K-4K}. As we can observe, the $F_1$-scores obtained by the nested ML model are similar to the ones obtained with a simple GAN, even after applying this filtering. Maybe, high values are obtained slightly more consistently with this filtering approach, but the results evidence that the improvement is not significative.

\subsection{Elitism by $F_1$-score}
\label{subsec:elit_f1}

Recall from subsection \ref{sec:results-standard-GAN} that with the generation of synthetic data through standard WGANs for each type of traffic, models were chosen randomly from all models obtained during the training phase. In this section, we shall explore a different strategy for sampling more efficiently the pair of models for data generation. Instead of picking a random model among all epochs, we will only draw samples from the top $10$ models obtained throughout all the training epochs, in the sense that they achieved the best $F_1$-scores. With this strategy, we aim to apply some type of elitism that prevents a drastic fall of the performance due to a random choice of a bad generator.
In addition, this strategy significantly reduces the number of combinations that we have to evaluate to find which pair of generators produces the best performing synthetic data.

For each type of traffic, we select the top $10$ models sorted by $F_1$-score. Drawing a sample uniformly at random from each top $10$ subset, we obtain a pair of generators with which we generate the synthetic dataset by following policy \ref{enum:poli-vainilla-1}), as described in subsection \ref{sec:results-standard-GAN}. The random selection of the pair of generators was done 20 times and the testing results obtained after using the synthetic data for training the Random Forest classifier are shown in Table \ref{tab:FA-linear-top10}. These results evidence that this strategy is very effective to increase the performance of the nested ML method. The obtained level of $F_1$-score outperforms the ones obtained with the standard sampling (Table \ref{tab:vainilla}), and are slightly below the obtained results with real data (Table \ref{table:real}) for the best choice of hyperparameter and even outperforms them with the default value. Additionally, the histograms plotted in Figure \ref{fig:vainilla-top10} evidence that with the elitism strategy, the results are much more consistent among executions and most of the results are around a $F_1$-score value of $0.95$ for any choice of the threshold.

Therefore, these results point out that this solution allows us to reach comparable results with fully synthetic datasets with respect to the ones obtained with real data. Moreover, the obtained data are robust, consistently leading to high values of $F_1$-scores. This also allows us to speed up the method of choosing the right generators. It is not necessary to conduct an exhaustive and long training with many epochs but, instead, it is better to save a small number of very good generators and to mix their results.
Furthermore, if the number of training epochs is sufficiently large, the exploration of the best combination of the pair of generators of the two traffic types can be limited to the evaluation of the subset combinations of the best top-K subsets of each traffic type.

\subsection{Elitism by statistical measures}
\label{sec:elit-metrics}

In this section, we shall explore a variant of the strategy used in the previous subsection \ref{subsec:elit_f1} for sampling with elitism. Again, we will sample the model used for generating the dataset from the top $10$ models in training. However, now, instead of using $F_1$-score as quality measure, we shall use the statistical measures described in Section \ref{sec:metrics} of $L^1$ distance and Jaccard index to sort the models. This analysis is useful to determine whether the statistical measures are a faithful quality control of the performance expected in a nested ML model.

As in subsection \ref{subsec:elit_f1}, policy \ref{enum:poli-vainilla-1} was used for generating the dataset. The results are shown in Table \ref{tab:FA-linear-top10-dKJ}. As the results evidence, the use of these statistical coefficients as quality measures leads to a substantial fall in the observed performance. 
This trend is also shown in the histograms plotted in Figure \ref{fig:vainilla-top10-DKJ}, where we observe that the $F_1$-scores obtained with this strategy are consistently lower than the ones reached with standard WGAN data (Figure \ref{fig:real_400K-4K} and Table \ref{tab:vainilla}) and with elitism by $F_1$-score (Figure \ref{fig:vainilla-top10} and Table \ref{tab:FA-linear-top10}).
It is worth noting that the results obtained with $L^1$ distance as a measure of quality tend to generalize better to the test split than the one obtained using Jaccard index. 
This is compatible with the observation that measuring the $L^1$ distance between the distributions is a much more complete comparison than just comparing their supports.

Consequently, the results of this section confirm the observation of subsection \ref{subsec:gan_vainilla_results}. Even though the theoretical results presented in the literature guarantee the convergence of the synthetic distribution to the original distribution, this convergence may not be correlated with a better performance of a nested ML model. 
Future works should explore new statistical metrics that better capture  the essence of the data that optimization algorithms utilise to train ML models.

\section{Conclusions and future work}
\label{sec:conclusions_future-work}

We propose a WGAN architecture to generate synthetic flow-based network traffic that can fully replace real traffic with two complementary goals: (1) avoiding privacy breaches when sharing data with third parties or deploying data augmentation solutions  and (2) obtaining a nearly unlimited source of synthetic data that is similar to the real data from a statistical perspective and can be utilised to fully substitute real data in ML training processes while keeping the same performance as ML models trained with real data.

To demonstrate the feasibility of our solution, we adopted a recently appeared cryptomining attack scenario in which two types of network traffic were considered: cryptomining attack connections and normal traffic consisting of web, video, P2P and email among others.
A set of four flow-based variables were selected to represent each connection in real-time. These variables have previously demonstrated their usefulness in ML-based cryptomining attack detectors and allow to compare the performance of GAN generators  with respect to naive approaches.

Instead of convolutional or recurrent networks, fully connected neural networks were used in WGANs architectures as no topological structure was present in the four selected features. Each type of traffic was modeled with two different WGAN architectures and using a rich set of hyperparameters we run an extensive number of WGAN trainings. Several enhancements were proposed to improve the simple WGAN performance: a custom activation function to better adapt the generator output to the data domain, and several heuristics to apply during GAN training to adapt the learning speeds of the generator and the discriminator.
Although the quality of the resultant synthetic data is similar when custom activation functions were applied in the output layer of the generator, the significant amount of non-existent synthetic data appearing in the simple GAN was almost entirely eliminated when these custom functions were used. We observed on a few occasions during  WGAN training that when the learning  process was temporarily blocked by one network, the {\em adaptive mini-batches} heuristic rapidly managed to rebalance the learning process. 
However, in our experiments, none of the other heuristics showed a significant effect on the quality of the synthetic data or the speed of convergence of the training process. Future work should investigate these heuristics more in depth to determine whether they can modulate or speed up the ill-convergence of the GAN training.

Due to the lack of metrics to measure the similarity of synthetic and real data in the network traffic domain, we defined two new metrics based on the $L^1$ distance and the Jaccard index to measure the quality of our synthetic data by comparing the join statistical distribution of synthetic and real data variables. 
Regarding the ill-convergence of GAN training, we propose a simple heuristic to be used as stopping criterion for GAN training. This heuristic selects the intermediate generator that produces the best performing synthetic data when used to train a ML-based cryptomining attack detector. In this context, the synthetic data performance metric is the $F_1$-score obtained by the ML-based attack detector in testing. For hyperparameter search, this heuristic is naturally extended to select the best WGAN configuration as the one with the largest $F_1$-score in any of its intermediate generators. 

It is worth noting that larger $F_1$-scores were obtained when flow-based variables of cryptomining connections were  generated than when normal traffic connections were replicated. We conjecture that the data distribution of normal connections is by far much complex as it is composed of many types of traffic (e.g., web, video, email, P2P) and on the contrary,  cryptomining connections although generated with four different protocols, share similar behaviour and therefore, their statistical patterns are much easier to replicate by WGANs. We increased the size of the WGAN architecture (layers and units) and the number of training epochs without being able to improve the $F_1$-score to the levels we achieved with the cryptomining WGAN. Future work should investigate how to break down such a complex data distribution into simpler distributions that can be easily replicated by WGANs.

Although in our experiments  decreases on $L^1$ and Jaccard metrics coincided with increases in the performance of synthetic data in ML tasks ($F_1$-score), we did not observe a strong correlation between the two types of metrics and therefore, we cannot apply the former instead of the latter, which entails greater computational costs. Future work should explore new computationally simple metrics that can accurately replace the costly evaluation of synthetic data performance in ML tasks we carried out during GAN training. 

We experimentally observed that 
using the best generator of two WGANs trained with different real data distributions to blend their synthetic data does not produce the best performance results when applied to the same ML task. 
Therefore, we propose for each type of data to select one generator uniformly at random from the set of intermediate generators obtained during the training of the GAN. Having obtained the blend of synthetic data to train the ML classifier, the $F_1$-score  is computed on testing. 
Elitism on $F_1$-score showed that the number of draws needed to achieve good performance decreased dramatically with respect to  pure random selection. Other elitisms based on $L^1$ distance and Jaccard index were experimented with, but the obtained results were not good and strengthen previous observations on the lack of strong correlation between these metrics and $F_1$-score. 
Future work should investigate why the combined synthetic data from the best performing generators of these two GANs do not produce the best performance when applied to the same ML task.

In addition to the previous open questions,
this manuscript points to several interesting challenges to be researched in future works:
\begin{itemize}
    \item Our custom activation functions only approximate exponential-like variable distributions using handcrafted LeakyRelu functions. Further research is needed to find activation functions that can fit any data distribution and in particular discrete and non-continous.
    
    \item Synthetic data generated by a GAN provides a way to circumvent real data privacy restrictions but a thorough and formal study on the reverse engineering of synthetic data should be conducted. 
    
    \item Ill-Convergence and oscillatory behaviour during GAN training is one of the key problems to be solved in GAN topic. Minimizing and maximizing partially in turns the cost function with respect to different variables tends to generate such oscillations and therefore, GAN optimization should be done in a more effective way.
    
    \item It is crucial to design computationally simple metrics that are strongly correlated with the performance of the synthetic data in ML tasks. These metrics could be used to drive the cost function during the GAN learning process, which would allow to implement an efficient stopping criterion.
    Furthermore, these metrics should be consistent when mixing synthetic data from two different distributions. Contrary to what we have observed in our experiments, it would be desirable that if the best generators from two different distributions are selected, the performance of the synthetic data obtained from the mixture should still be the best when the combined synthetic data is applied to the same ML task.
    
    \item The synthetically generated flow-based variables represent the state of a connection at a specific instant in time. It would be interesting to develop new GAN architectures to generate synthetic time series of these variables for use in the training of more complex IDS.
\end{itemize}

\section*{Acknowledgements}
This work was partially supported by the European Union’s Horizon 2020 Research and Innovation Programme under Grant 833685 (SPIDER) and Grant 101015857 (Teraflow).
The second author acknowledges the hospitality of the Department of Mathematics at Universidad Aut\'onoma de Madrid, where part of this work was conducted. The second author has been partially supported by Spanish Ministerio de Ciencia e Innovaci\'on through project PID2019-106493RB-I00 (DL-CEMG).


\clearpage
\begin{appendices}

\section*{Appendix I: Figures}
\label{sec:Appendix}

\begin{figure*}[!h]
\begin{mdframed}
\centering

\begin{subfigure}[t]{0.99\textwidth}
\includegraphics[width=0.49\linewidth]{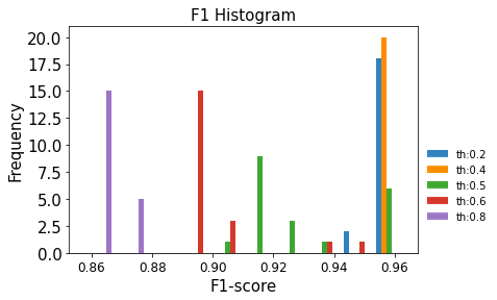} 
\includegraphics[width=0.49\linewidth]{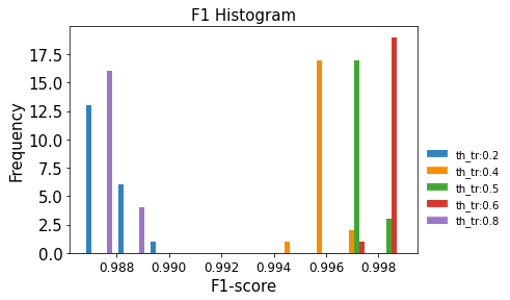}
\caption{F1-score on testing (left) and training (right) using real data for training and testing (400K/4K distribution).}\label{fig:real_400K-4K}
\end{subfigure}

\medskip

\begin{subfigure}[t]{.99\textwidth}
\centering
\vspace{0pt}
\includegraphics[width=0.49\linewidth]{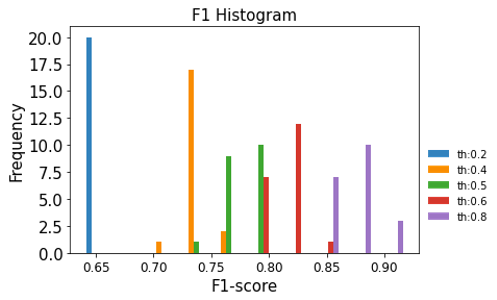} \includegraphics[width=0.49\linewidth]{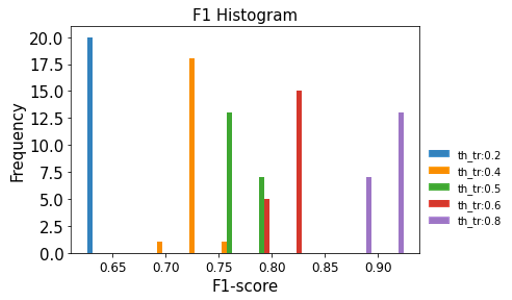}
\caption{F1-score on testing (left) and training (right) using real data for training and testing (4K/4K distribution).}\label{fig:real_4K-4K}
\end{subfigure}
\end{mdframed}
\medskip
\begin{minipage}[t]{.99\textwidth}
\caption{$F_1$-score on testing (left) and training (right) using real data  with 400K/4K (\subref{fig:real_400K-4K}) and 4K/4K (\subref{fig:real_4K-4K}) class distributions for training (subsection \ref{sec:experiments-real}). Results for decision thresholds of 0.2, 0.4, 0.5, 0.6 and 0.8 are represented.} \label{fig:real}
\end{minipage}
\end{figure*}

\begin{figure*}[!tb]
\begin{mdframed}
\centering

\begin{subfigure}[t]{0.99\textwidth}
\centering
\includegraphics[width=0.49\linewidth]{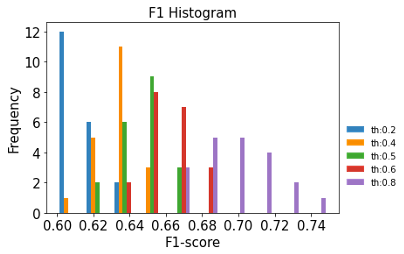} \includegraphics[width=0.49\linewidth]{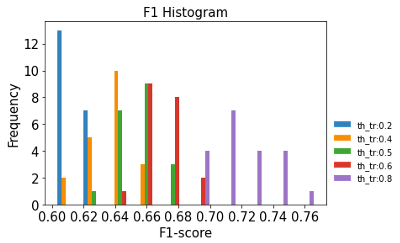}
\caption{Unbalanced dataset with the original 400K/4K distribution.}\label{fig:mean_400K-4K}
\end{subfigure}

\medskip

\begin{subfigure}[t]{.99\textwidth}
\centering
\vspace{0pt}
\includegraphics[width=0.49\linewidth]{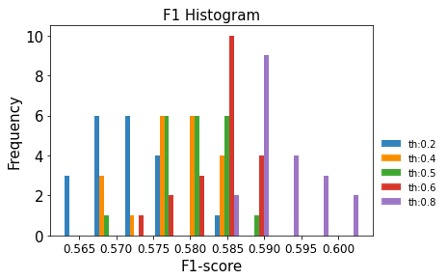} \includegraphics[width=0.49\linewidth]{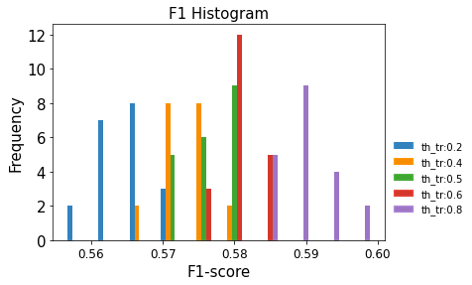}
\caption{Balanced dataset with 4K/4K distribution.}\label{fig:mean_4K-4K}
\end{subfigure}
\end{mdframed}

\medskip
\begin{minipage}[b]{.99\textwidth}
\caption{$F_1$-score on testing (left) and training (right) using a na\"ive mean-based generator with unbalanced and balanced datasets for training (subsection \ref{sec:naive-exp}). Results for decision thresholds of 0.2, 0.4, 0.5, 0.6 and 0.8 are represented. 
} \label{fig:mean}
\end{minipage}

\end{figure*}

\begin{figure*}[!hp]
\begin{mdframed}
\centering

\begin{subfigure}[t]{0.99\textwidth}
\centerline{
  \includegraphics[width=0.49\linewidth]{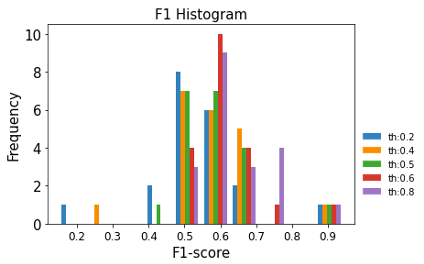} 
  \includegraphics[width=0.49\linewidth]{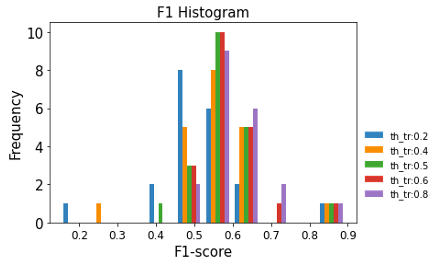}}
\caption{Policy \ref{enum:poli-vainilla-1}). Training with 400K/4K distribution and one generator chosen uniformly at random}\label{fig:vainilla_400K-4K}
\end{subfigure}
\medskip

\begin{subfigure}[t]{.99\textwidth}
\centering
\vspace{0pt}
\includegraphics[width=0.49\linewidth]{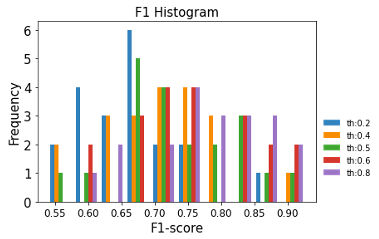}
\hfill
\includegraphics[width=0.49\linewidth]{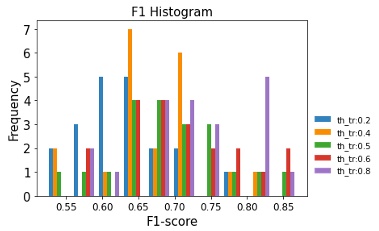}
\caption{Policy \ref{enum:poli-vainilla-2}). Training with 400K/4K distribution and a mix of two generators is chosen uniformly at random}\label{fig:vainilla2_400K-4K}
\end{subfigure}

\medskip
\begin{subfigure}[t]{0.99\textwidth}
\centerline{
  \includegraphics[width=0.49\linewidth]{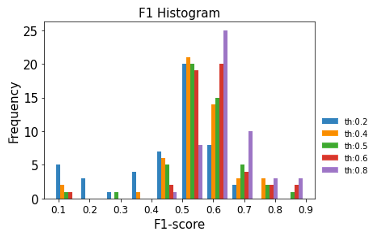} 
  \includegraphics[width=0.49\linewidth]{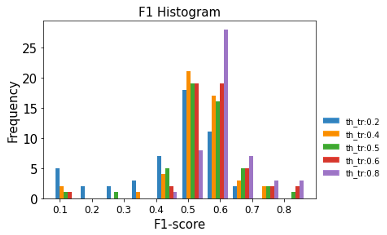}}
\caption{Policy \ref{enum:poli-vainilla-4}). Training with 4K/4K distribution and one generator chosen uniformly at random}\label{fig:vainilla_4K-4K}
\end{subfigure}
\end{mdframed}
\medskip
\begin{minipage}[t]{.99\textwidth}
\caption{$F_1$-score on testing (left) and training (right) using a standard GAN generator and sampling policies \ref{enum:poli-vainilla-1}), \ref{enum:poli-vainilla-2}) and \ref{enum:poli-vainilla-4}) (subsection \ref{sec:results-standard-GAN}). Results for decision thresholds of 0.2, 0.4, 0.5, 0.6 and 0.8 are represented.} \label{fig:vainilla}
\end{minipage}

\end{figure*}
\begin{figure*}[!hp]
\begin{mdframed}
\centering

\begin{subfigure}[t]{0.99\textwidth}
\centerline{
  \includegraphics[width=0.49\linewidth]{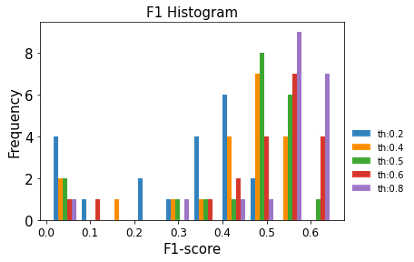} 
  \includegraphics[width=0.49\linewidth]{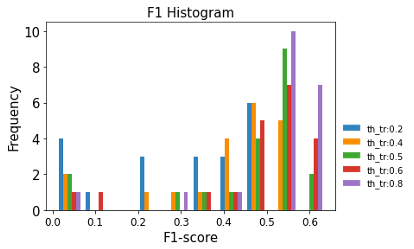}}
\caption{Policy \ref{enum:poli-vainilla-1}). Training with 400K/4K distribution and one generator chosen uniformly at random}\label{fig:FA_400K-4K}
\end{subfigure}

\medskip

\begin{subfigure}[t]{.99\textwidth}
\centering
\vspace{0pt}
\includegraphics[width=0.49\linewidth]{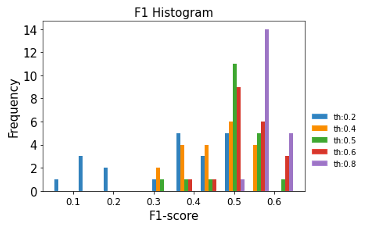}
\hfill
\includegraphics[width=0.49\linewidth]{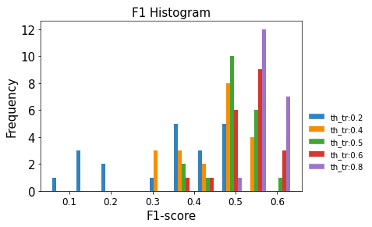}
\caption{Policy \ref{enum:poli-vainilla-2}). Training with 400K/4K distribution and a mix of two generators is chosen uniformly at random}\label{fig:FA-2_400K-4K}
\end{subfigure}

\medskip
\begin{subfigure}[t]{0.99\textwidth}
\centerline{
  \includegraphics[width=0.49\linewidth]{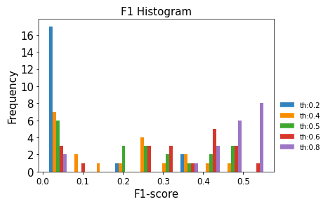} 
  \includegraphics[width=0.49\linewidth]{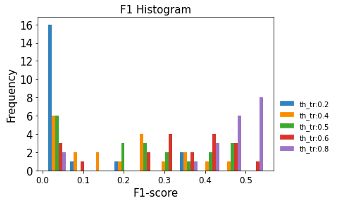}}
\caption{Policy \ref{enum:poli-vainilla-4}). Training with 4K/4K distribution and one generator chosen uniformly at random}\label{fig:FA_4K-4K}
\end{subfigure}
\end{mdframed}
\medskip
\begin{minipage}[t]{.99\textwidth}
\caption{$F_1$-score on testing (left) and training (right) using a generator with custom activation functions at the output and policies \ref{enum:poli-vainilla-1}), \ref{enum:poli-vainilla-2}) and \ref{enum:poli-vainilla-4}) (subsection \ref{sec:custom-act-func}). Results for decision thresholds of 0.2, 0.4, 0.5, 0.6 and 0.8 are represented.} \label{fig:FA}
\end{minipage}

\end{figure*}

\begin{figure*}[!h]
\begin{mdframed}
\centering

\begin{subfigure}[t]{0.99\textwidth}
\centerline{
  \includegraphics[width=0.49\linewidth]{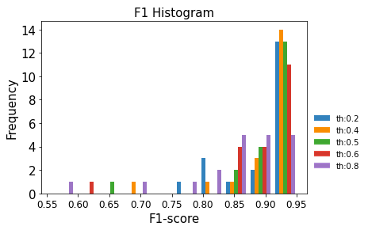} 
  \includegraphics[width=0.49\linewidth]{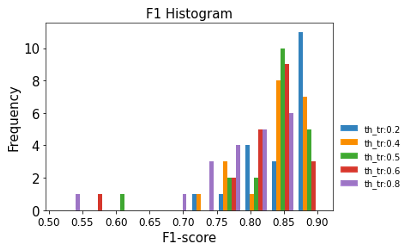}}
\caption{Training with 400K/4K distribution. 1 generator chosen uniformly at random}\label{fig:vainilla-top10_400K-4K}
\end{subfigure}
\medskip

\begin{subfigure}[t]{.99\textwidth}
\centering
\vspace{0pt}
\includegraphics[width=0.49\linewidth]{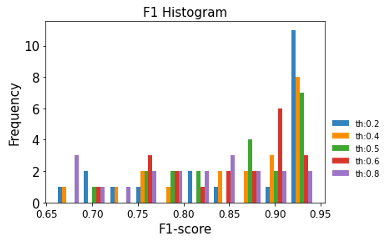}
\hfill
\includegraphics[width=0.49\linewidth]{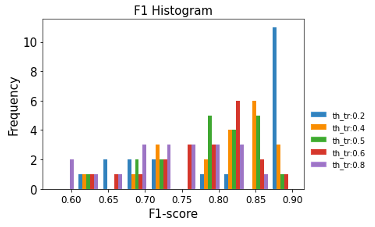}
\caption{Training with 400K/4K distribution. 1 generator chosen uniformly at random} filtering positive values\label{fig:vainilla-Pos-top10_400K-4K}
\end{subfigure}

\medskip
\begin{subfigure}[t]{0.99\textwidth}
\centerline{
  \includegraphics[width=0.49\linewidth]{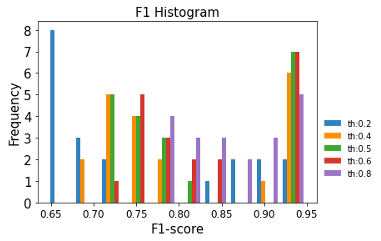} 
  \includegraphics[width=0.49\linewidth]{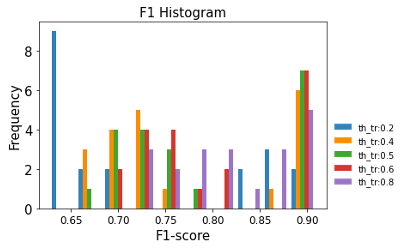}}
\caption{Training with 4K/4K distribution. 1 generator chosen uniformly at random}\label{fig:vainilla-top10_4K-4K}
\end{subfigure}
\end{mdframed}
\medskip
\begin{minipage}[t]{.99\textwidth}
\caption{$F_1$-score on testing (left) and training (right) with sampling elitism among the top $10$ models in training sorted by $F_1$-score (subsection \ref{subsec:elit_f1}). Results for decision thresholds of 0.2, 0.4, 0.5, 0.6 and 0.8 are represented.} \label{fig:vainilla-top10}
\end{minipage}

\end{figure*}
\begin{figure*}[!htp]
\begin{mdframed}
\begin{subfigure}[t]{1.\textwidth}
\centerline{
    \includegraphics[width=0.49\linewidth]{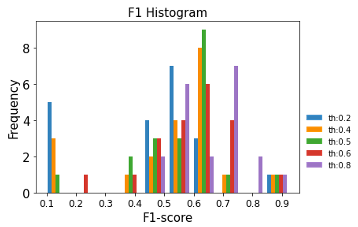} 
    \includegraphics[width=0.49\linewidth]{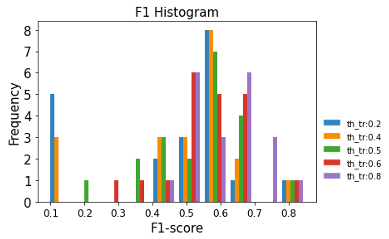}}

\end{subfigure}

\begin{minipage}[t]{1.\textwidth}
\caption{$F1-score$ on testing (left) and training (right) using the discriminator as a quality assurance filter (subsection  \ref{sec:discr-quality-assurance}). Training with 400K/4K distribution. Results for decision thresholds of 0.2, 0.4, 0.5, 0.6 and 0.8 are represented.}
\label{fig:vainillaPos_400K-4K}
\end{minipage}
\end{mdframed}
\end{figure*}

\begin{figure*}[!h]
\begin{mdframed}
\centering

\begin{subfigure}[t]{0.99\textwidth}
\centerline{
  \includegraphics[width=0.49\linewidth]{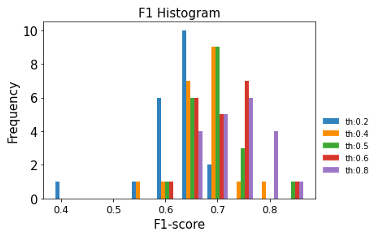} 
  \includegraphics[width=0.49\linewidth]{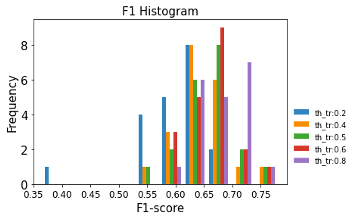}}
\caption{Using top $10$ models sorted by $L^1$ distance}
\end{subfigure}
\medskip

\begin{subfigure}[t]{.99\textwidth}
\centering
\vspace{0pt}
\includegraphics[width=0.49\linewidth]{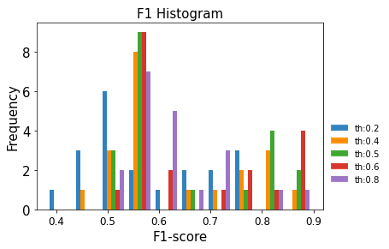}
\hfill
\includegraphics[width=0.49\linewidth]{images/Vainilla_top10_Pos_1rnd_0-1_400-4_Training.png}
\caption{Using top $10$ models sorted by Jaccard index} 
\end{subfigure}

\end{mdframed}
\medskip
\begin{minipage}[t]{.99\textwidth}
\caption{
$F1-score$ on testing (left) and training (right) with sampling elitism using policy \ref{enum:poli-vainilla-1}). Elitism of the top $10$ sorted by statistical coefficients (subsection \ref{sec:elit-metrics}. Results for decision thresholds of 0.2, 0.4, 0.5, 0.6 and 0.8 are represented.} \label{fig:vainilla-top10-DKJ}
\end{minipage}

\end{figure*}

\end{appendices}

\end{document}